\documentclass[12pt, a4paper]{article}

\usepackage[
  margin=0.75in,
  headsep=10pt, 
]{geometry}

\usepackage{graphicx}
\usepackage{soul}
\usepackage{graphicx}
\usepackage{epstopdf, epsfig}
\usepackage{amsmath}
\usepackage[table]{xcolor}
\usepackage{subcaption}
\usepackage{soul}
\usepackage{makecell}
\usepackage{multirow}
\usepackage{breqn}
\usepackage{amssymb} 
\usepackage{booktabs}
\usepackage{dcolumn}
\usepackage{bm}
\usepackage[utf8]{inputenc}
\usepackage[T1]{fontenc}
\usepackage{mathptmx}
\usepackage{etoolbox}
\usepackage{hyperref}
\usepackage{subcaption}
\usepackage{ragged2e}
\usepackage[sort&compress]{natbib} 
\usepackage[utf8]{inputenc}
\usepackage[T1]{fontenc}
\usepackage{graphicx}
\usepackage{epstopdf, epsfig}
\usepackage{xcolor}
\usepackage[table]{xcolor}
\usepackage{soul}
\usepackage{makecell}
\usepackage{multirow}
\usepackage{booktabs}
\usepackage{dcolumn}

\newcommand{\edt}[1]{{\color{black}#1}} 
\newcommand{\dn}[1]{{\color{black}#1}} 

\usepackage{amsmath}   
\usepackage{amssymb}   
\usepackage[normalem]{ulem}

\newcommand{\soptitle}{Wing-Rotor Aerodynamic Interactions in Small UAVs During Hover and Cruise}
\usepackage{xcolor}

\begin{document}

\begin{center}
\Large \bf{\soptitle}
\vspace{0.1in}
\end{center}

\begin{center}
{Dev Pradeepkumar Nayak$^1$, Seungmin Choi$^{1,2}$, and Muhammad Saif Ullah Khalid$^{1,\ast}$}
\vspace{0.1in}
\end{center}
\begin{center}
$^1$Nature-Inspired Engineering Research Lab (NIERL), Department of Mechanical and Mechatronics Engineering, Lakehead University, Thunder Bay, ON P7B 5E1, Canada\\
$^2$Department of Unmanned Aircraft Systems, Hanseo University, Seosan, Republic of Korea
\vspace{0.05in}
$^\ast$\small{Corresponding Author, Email: mkhalid7@lakeheadu.ca}
\end{center} 

\begin{abstract}
A compact vertical take-off and landing aircraft requires the same tilt-rotor configuration to perform two fundamentally different aerodynamic tasks: sustain hover and deliver efficient cruise. This \edt{work} investigates the underlying wing-rotor interaction\edt{s} in both operating regimes using a validated unsteady Reynolds-averaged Navier-Stokes equations\edt{-based computational} framework. \edt{For} cruise, the advance ratio governs the balance between thrust production, propulsive efficiency, and wake coherence. Lower advance ratios produce a tightly wound slipstream that undergoes strong vortex interactions, leapfrogging, and early wake bifurcation. At higher advance ratios, the slipstream retains a narrower and more coherent jet-like structure, improving propulsive efficiency while reducing both thrust \edt{of the propellor} and lift \edt{of the wing}. The flow impingement at the wing is characterized through the approaching, interaction, and convection phases, revealing the combined influence of vortex stretching, wake bifurcation, blockage, image-induced velocity, and streamwise momentum convection on the downstream wake. In hover, \edt{the} propeller\edt{'s} rotational speed governs the overall aerodynamic performance more strongly than \edt{the} wing\edt{'s} placement. Although the \edt{position of the wing with respect to the propellor} modifies the local wake interaction\edt{s} and flow impingement \edt{on its leading edge}, its influence on the integrated thrust coefficient and figure of merit remains limited. We also explain the vortex and wake dynamics around the propellor and the wing responsible for governing these aerodynamic performance. 
\end{abstract}

\section{INTRODUCTION}


Recent advancements in drone technology \edt{facilitates} their application across a broad spectrum of fields, ranging from recreational activities and high-end cinematography to sophisticated military operations \citep{hassanalian2017classifications}. These developments \edt{are} driven by substantial progress in engineering methodologies, particularly in the advancement of reliable computational and experimental fluid dynamics frameworks \citep{abdulrahman2025review}. Broadly, \edt{presently existing} \edt{u}nmanned \edt{a}erial \edt{v}ehicles (UAVs) can be categorized into two primary \edt{classes, including} fixed-wing UAVs and multi-rotor UAVs. Fixed-wing UAVs resemble scaled-down aircraft, generating lift through forward motion, whereas multi-rotor UAVs operate similarly to helicopters, producing lift directly through rotating propellers, enabling vertical takeoff and hovering capabilities \citep{hassanalian2017classifications, zawodny2021aerodynamic, bailly2023jaxa, abdulrahman2025review}.


Both categories offer distinct advantages while also exhibiting \edt{some} inherent limitations \edt{too}. To overcome these constraints, a third class of UAVs emerged, commonly referred to as vertical take-off and landing (VTOL) aircraft \citep{zawodny2021aerodynamic, simmons2021system}. These configurations combine the capabilities of multi-rotor and fixed-wing aircraft, enabling vertical takeoff and landing through a rotor-supported flight mode, followed by a transition to forward flight through \edt{tilting rotors. Therefore,} their operation can be broadly divided into three primary flight states: \edt{(i)} cruise, corresponding to forward flight, and \edt{(ii)} hover\edt{, relating to holding a station aerially, and (iii) vertical} takeoff and landing. Although the applications and parametric scaling of such UAVs \edt{were} extensively investigated, studies focusing on smaller-scale configurations\edt{, specifically making it relevant to low Reynolds numbers remained} comparatively limited. UAVs classified as ``mini'' are generally characterized by a representative dimension \edt{related to the propellor with a diameter} of less than $\dn{0.1~\mbox{m}}$ \dn{for its blades} \citep{hassanalian2017classifications}. Accordingly, the configuration considered in \edt{our}{ present investigation may be classified as a mini-UAV, with a propeller\edt{'s} diameter of $D_p=0.1~\mbox{m}$.

Owing to the close proximity of the wing and propeller \edt{VTOL aerial vehicles}, it is crucial to investigate the design and operating parameters that govern the aerodynamic performance of the \edt{UAV} at different \edt{cruising} speeds and during hover or takeoff\edt{/landing}. Within the complex framework of a VTOL UAV transitioning between hover and cruise, design or operating parameters that improve performance in one flight regime may have a detrimental effect in the other, and vice versa. Therefore, both the design parameters and operating parameters must be selected with consideration of their coupled effects, such that an appropriate balance is achieved between the overall hover and cruise performance of the VTOL UAV.

For the VTOL UAV operating under cruise conditions, the placement of the wing near the propeller wake can be considered analogous to the configuration investigated by Muscari et al. \citep{muscari2017analysis}. Although their study focused on a marine propeller, the underlying principles governing the flow field, wake development, and vortex interactions between the propeller blades and the wing remain relevant to the \edt{presently considered} configuration. Previous studies examining wing–rotor interactions in forward flight primarily considered the advance ratio, \edt{cruising} velocity, and relative placement \edt{of the wings} as the major operating and design parameters \citep{deters2014reynolds, baek2015effects, felli2011mechanisms, felli2021underlying, muscari2017analysis}. Accordingly, the primary objective of the present study under \edt{cruising} conditions is to identify the combination of these parameters that yields the highest propulsive efficiency. The resulting quantitative trends are further interpreted through the analysis of vortex–body and vortex–vortex interactions. Under cruise conditions, the wing\edt{'s} position is fixed at \edt{a distance of} $0.5R$ from the nacelle and $0.5R$ from the propeller-blade centreline\edt{. These specifications come from} the configuration investigated by Zawodny \textit{et al.} \cite{zawodny2021aerodynamic}. Their study examined the influence of \edt{the} wing\edt{'s} placement and demonstrated that positioning the wing closer to both the propeller and the nacelle resulted in higher thrust generation. This finding provides the primary justification for maintaining the position \edt{of the wing} unchanged throughout \edt{our simulations related to cruise flight}.

For the cruise conditions \edt{considered in our current work}, classical scaling\edt{-related} arguments indicate that, to maintain \edt{levels of} thrust comparable to those of larger-scale propellers, a small-scale propeller must operate at a higher rotational speed, which consequently alters the resulting wake structure \citep{deters2014reynolds}. Based on the operating ranges commonly reported in the literature, the propeller is expected to perform \edt{well} over a broad range of advance ratios ($J = 0.1-1.0$), and at \edt{cruising speeds} ($U_{\infty}$) of up to $30~\mbox{m/s}$ \citep{deters2014reynolds,baek2015effects,felli2011mechanisms}. \edt{Here, we define the advance ratio as $J={U_\infty}/{n}{D_p}$, where $n$ denotes the angular frequency of the propellor, and $D_p$ is diameter of the propellor.}

Under hover\edt{ing} conditions, the absence of a freestream requires the propeller to generate enough momentum needed to support the vehicle, \edt{which makes} induced power a dominant aerodynamic requirement \citep{strawn2002computational,cameron2016performance}. The resulting wake \edt{gets contracted} downstream and contains concentrated helical tip vortices whose trajectory, circulation, and core development strongly influence \edt{the} induced velocity, thrust, torque, and \edt{loading on the} blade \citep{young2003vortex,okulov2007stability}. When a wing is located within this wake, vortex impingement and accelerated downwash can produce non-uniform loading and reduce the net vertical force available to the aircraft \citep{felker1988aerodynamic, young2002rotorwing}. These effects depend mainly on \edt{the} propeller\edt{'s} speed and relative placement \edt{of the wing, making it an important aspect of operating VTOL UAVs and one of the main research objectives of our present study}. 

\edt{With this context and background, we particularly focus on addressing the following important yet unanswered research questions: (i) \dn{how does the propeller-wing interactions affect the aerodynamic performance across the range of conventional advance ratios in cruise? (ii) how does the positioning of the wing affect the performance of the propeller over a wide range of operating angular speeds in hover? and (iii) how does the presence of the wing influence the evolution of the propeller's slipstream in both cruise and hover? Addressing these aspects is important to analyze the performance of a tilting rotor configuration under both cruise and hover conditions to optimize its overall aerodynamic capabilities in VTOL UAVs}. To define a more focused yet representative parametric space, the present investigation considers freestream velocities of $4\edt{~\mbox{m/s}}$, $8\edt{~\mbox{m/s}}$, and $12~\mbox{m/s}$, with the advance ratio \edt{ranging} from $J = 0.2-0.8$ \edt{of propellors} for the investigations \edt{of propellor-wing interactions during} cruise \edt{conditions}. \edt{Nevertheless, there is no freestream aerial flow for} hover\edt{-related} conditions\edt{. Here, we consider} a broad range of} rotational speeds \edt{of the propellor}, comprising $8,900\edt{~\mathrm{RPM}}$, $11,430\edt{~\mathrm{RPM}}$, $15,740\edt{~\mathrm{RPM}}$, and $24,200~\mathrm{RPM}$. \edt{We also consider specific locations for placement of the wing with respect to the propellor,} \dn{as shown in Fig.~\ref{fig:wingplacement}}.



\section{COMPUTATIONAL METHODOLOGY}

\edt{In this research work, we consider} the Gemfan $51466$ propeller, which is commonly used in first-person view ($\mathrm{FPV}$) drones \edt{\citep{kim2025rapid}}. The present work focuses on analyzing the operation of a VTOL configuration at high angular velocities. Although airfoils\edt{,} such as NACA $4424$ can be effective for lift generation, they typically exhibit higher drag coefficients \edt{\citep{panagiotopoulos2024design}}. \edt{Contrarily}, more streamlined symmetric airfoils produce a uniform pressure distribution between the upper and lower surfaces, and may not generate sufficient lift \edt{\citep{panagiotopoulos2024design}}. \edt{Here, we select} NACA $4412$ airfoil for the wing\edt{'s} profile. This airfoil offers a favorable balance between thickness and camber, enabling the generation of sufficient pressure difference for lift production while maintaining relatively low drag. The main geometrical features of the propeller are reported in Table.~\ref{tab:propeller}.

\begin{table}[h!]
\centering
\caption{\edt{Geometric parameters for the propeller}}
\label{tab:propeller}
\begin{tabular}{lc}
\hline
\hline
\multicolumn{2}{c}{Gemfan 51466 FPV model} \\
\hline
Radius & $R = 0.05~\dn{m}$ \\
Number of blades & $Z = 3$ \\
Pitch ratio & $P/D = 0.90$ \\
Hub ratio & $0.12$ \\
\hline
\hline
\end{tabular}
\end{table}

The VTOL configuration is illustrated in Fig.~\ref{fig:schematic}, along with the non-dimensional propeller radius and wing chord length. The schematics presented in Figs.~\ref{fig:schematic}a and \ref{fig:schematic}b represent the orientations of the propeller-wing configuration with respect to an aircraft (outlined in blue) in cruise and hover, respectively. \edt{Please note that} this outline of \edt{an airplane} is included solely to indicate the orientation of the VTOL flight and \edt{is} not \edt{part of the currently reported simulations}.). The normalized locations of the wing relative to the centerline \edt{of the propellor} and nacelle are shown in the enlarged views in Figs.~\ref{fig:schematic}c and \ref{fig:schematic}e for the cruise and hover configurations, respectively. Figure~\ref{fig:schematic}d shows an isometric view of the VTOL configuration and identifies the components considered in \edt{this work}. In Figs.~\ref{fig:schematic}c and \ref{fig:schematic}e, $\Delta X$ and $\Delta Y$ represent the axial and vertical locations of the wing leading edge relative to the propeller centerline and nacelle, respectively.

\begin{figure}[h!]
    \centering
    \includegraphics[width=1\linewidth]{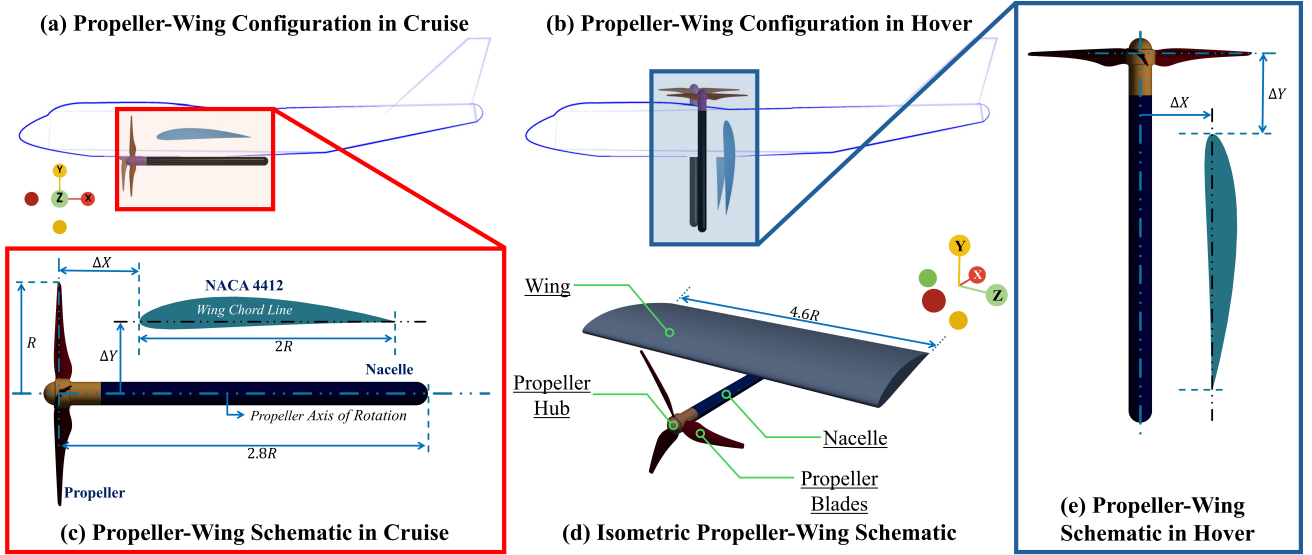}
    \caption{Propeller–wing configuration in (a) cruise and (b) hover. Detailed views illustrate the corresponding orientations in (c) cruise, (d) isometric view, and (e) hover. (\textit{Note:} The aircraft outline is included solely to indicate the orientation of the VTOL flight and \edt{is not part of the present simulations}.)}
    \label{fig:schematic}
\end{figure}

We perform our computational simulations in OpenFOAM-v2412 \edt{by solving} unsteady Reynolds-averaged Navier-Stokes (URANS) \edt{equations}, employing $\text{k–}\omega$ Shear Stress Transport (SST) turbulence model \edt{\citep{menter1994two}}. This hybrid model combines the near-wall accuracy of the $k$--$\omega$ model and the free-stream robustness of the $\text{k–}\epsilon$ model \edt{\citep{launder1983numerical, younoussi2024calibration}}. The SST model incorporates a shear stress limiter to improve predictions under adverse pressure gradients and flow separation. The non-dimensional continuity and incompressible URANS turbulence equations are given below \cite{menter1994two, younoussi2024calibration, launder1983numerical, dave2020variable}.

\begin{equation}
\nabla \cdot \bar{\mathbf{u}} = 0
\label{eq:nondim_continuity}
\end{equation}

\begin{equation}
\frac{\partial \bar{\mathbf{u}}}{\partial t}
+ \bar{\mathbf{u}} \cdot \left({\nabla} \bar{\mathbf{u}}\right)
=
-\nabla \bar{p}
+\frac{1}{Re}\nabla^{2}\bar{\mathbf{u}}
-\nabla \cdot \tau
\label{eq:nondim_momentum}
\end{equation}

\noindent Where $p$ is the pressure, and the last term, $\tau_{ij} = \overline{u_i' u_j'}$, corresponds to the Reynolds stress tensor, which accounts for the transport of momentum induced by turbulent velocity fluctuations. \edt{For closure of the governing set of equations}, Reynolds stresses are modeled using the Boussinesq eddy-viscosity approximation, relating them to the mean strain rate through an eddy viscosity. In Eqs.~\ref{eq:nondim_continuity} and \ref{eq:nondim_momentum}, $\bar{\mathbf{u}}$ denotes the normalized mean velocity vector, $\bar{p}$ is the normalized pressure, and $\mbox{Re}$ is the Reynolds number \edt{for} characterizing the flow. The temporal term is discretized using an implicit backward difference scheme. This algorithm combines the Pressure-Implicit with Splitting of Operators ($PISO$) algorithm and Semi-Implicit Method for Pressure-Linked Equations ($SIMPLE$) algorithm. The convergence criterion for the iterative solution at each time step is set to $10^{-04}$ \edt{\citep{jasak2009dynamic}}.

\begin{figure}[h!]
    \centering
    \includegraphics[width=1\linewidth]{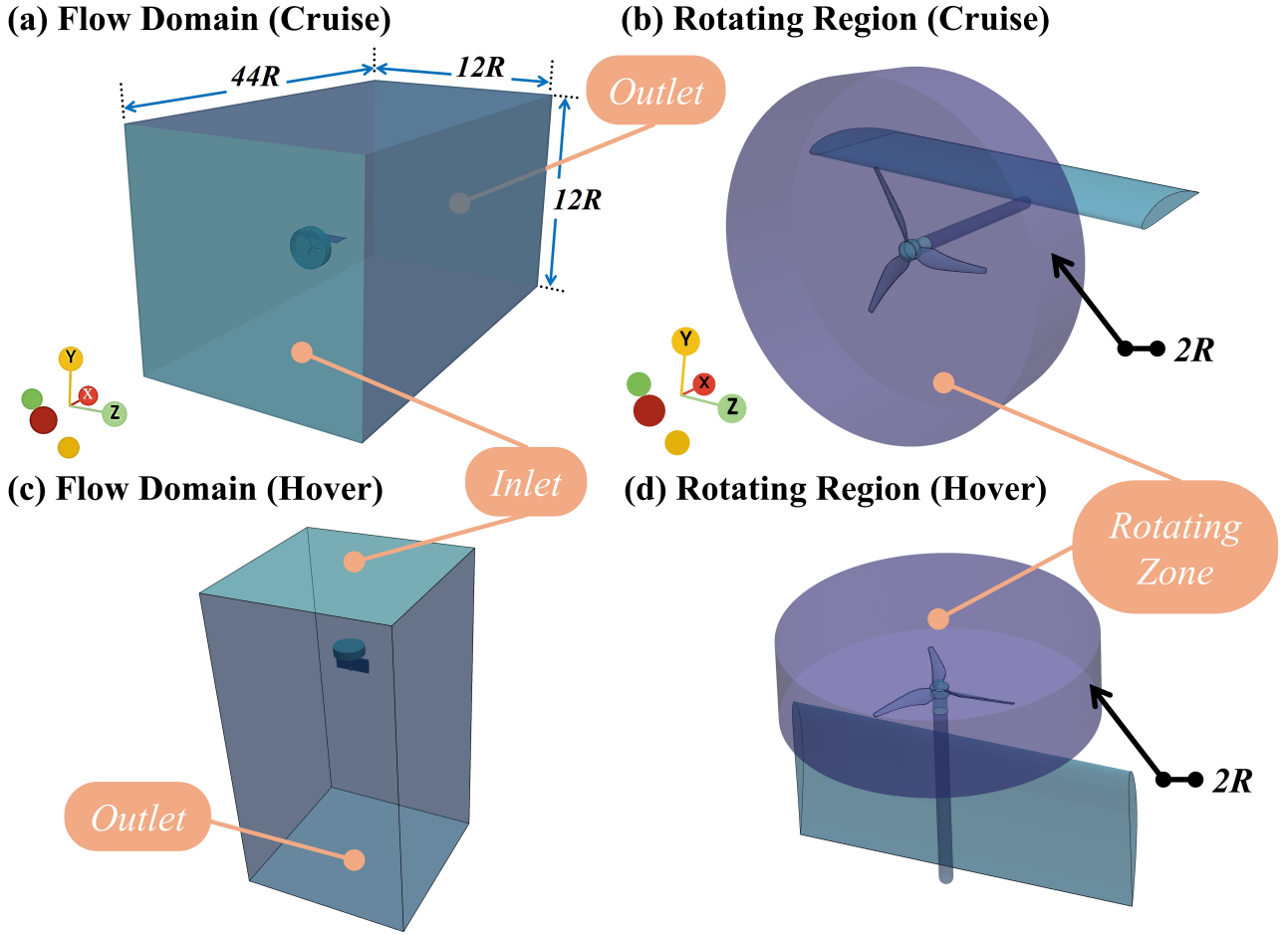}
    \caption{Computational domain and rotating region. (a \edt{\&} c) static zones for the propeller in cruise and hover, respectively, normalized by the propeller\edt{'s} radius $R$\edt{,} (b \edt{\&} d) rotating zones of radius $2R$ enclosing the propeller in cruise and hover, respectively}
    \label{fig:domain}
\end{figure}

Figure~\ref{fig:domain} shows the unstructured grid used \edt{for our simulations}. The domain in Fig.~\ref{fig:domain}a and \ref{fig:domain}c is normalized \edt{using} the radius of the propeller. \edt{In order} to handle the rotational motion of the propeller, a sliding mesh technique is used. The mesh is segregated in two regions, \edt{including a} static zone \edt{that} belongs to the part of the main domain, in which the rotating zone of the mesh is enclosed\edt{,} as shown in Fig.~\ref{fig:domain}b for cruise and Fig.~\ref{fig:domain}d for hover. The rotating zone encloses the hub and the propeller blades and has a radius of $2R$. The interpolation \edt{of numerics} between the static and the rotating zone is handled using a cyclic $\mbox{AMI}$-type boundary condition, and the tolerance is set to $1e-04$ \edt{\citep{jasak2009dynamic}}.

\begin{figure}[h!]
    \centering
    \includegraphics[width=1\linewidth]{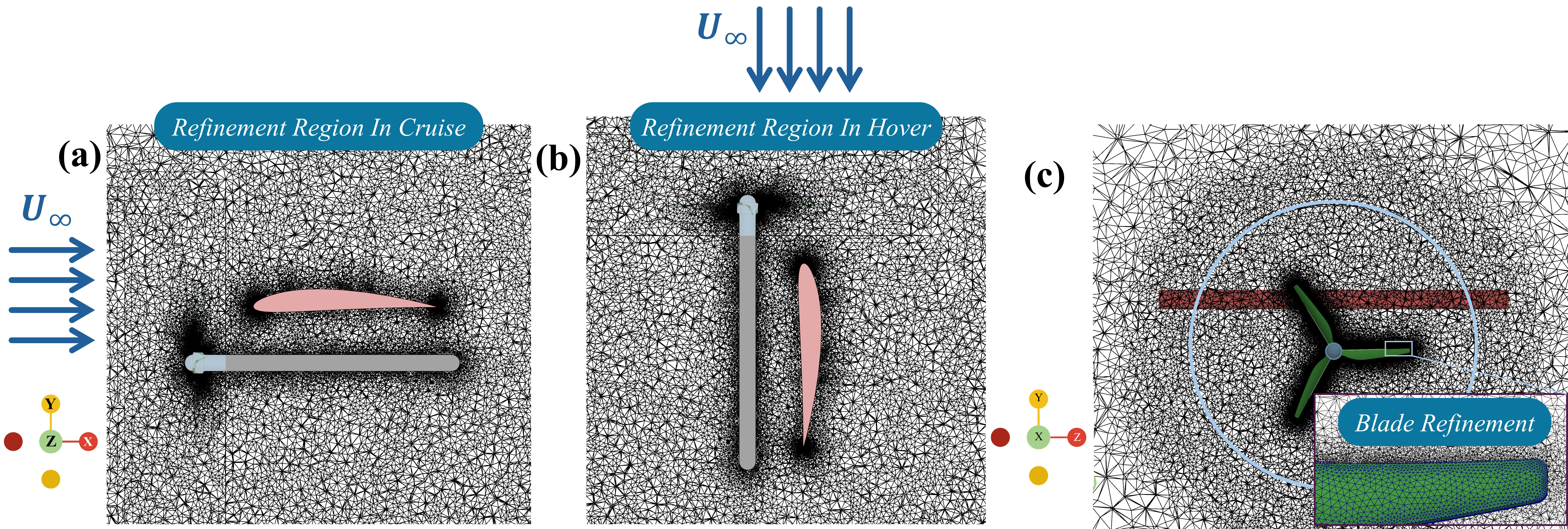}
    \caption{Computational mesh and refinement \edt{levels, where the} global mesh \edt{has} refinement around the wing and nacelle for the propeller in (a) cruise, and (b) hover, \edt{and} (c) local refinement \edt{is} within the rotating zone, highlighting \edt{highly} resolved mesh near the blade of the propeller.}
    \label{fig:mesh}
\end{figure}

In Fig.~\ref{fig:mesh}, the grid refinement around the geometry is shown from sectional and frontal views. The near-wall regions are resolved with a maximum $y^+$ value of approximately $10$, corresponding to the buffer layer of wall-bounded turbulent flow. This region lies between the viscous sublayer and the logarithmic layer, where neither a fully wall-resolved nor a conventional wall-function treatment is strictly applicable. However, these intermediate $y^+$ values \edt{are} retained in \edt{our} URANS simulations employing the $\text{k-}\omega$ SST model to balance computational cost and near-wall resolution. Around \edt{the} curved surfaces, the $y^+$ value decreases further toward $1$. The refined grids around the propeller, nacelle, and wing are shown in Figs.~\ref{fig:mesh}a, \ref{fig:mesh}b, and \ref{fig:mesh}c, respectively. Figures~\ref{fig:mesh}a and \ref{fig:mesh}b also \edt{exhibit} \edt{directions of} the inlet \edt{flow} for the cruise and hover configurations. As stated previously, $U_{\infty}=0$ \edt{for the hovering configurations}.

\subsection{Verification \& validation}

The grid is resolved enough to ensure the validity of the $\text{k–}\omega$ SST model\edt{. To} further \edt{justify resolution of the} grid and confidence in \edt{in our meshing approach}, the grid\edt{-convergence tests are} performed \edt{by choosing} three grid resolutions. For the grid-convergence and time-step \edt{independence tests}, we perform the simulation\edt{s} for $20$ total revolutions, with $J=0.7$ and $U_\infty=1$. The simulation\edt{s} attain steady state \edt{solutions} within $8$ revolutions. The three grids chosen for the grid-convergence study are: Grid~01 with 19,315,706 ($19M$) cells, Grid~02 with 39,772,704 ($39M$) cells, and Grid~03 with 51,544,856 ($52M$) cells. Figure~\ref{fig:gridInd} shows the wing-tip vortex and the bound vortex generated around the wing, along with the tip vortices shed from the blades of the propeller. Upon closer inspection of the highlighted region \edt{of a blade of the propellor} in \edt{Figs}.~\ref{fig:gridInd}a–\ref{fig:gridInd}c, \edt{we} observed that \edt{an} increasing \edt{refinement of}, of the grid \edt{helps} capture the vortex formation and detachment from the surface of the blades. With finer grids, the near-blade vortical structures appear more coherent and resolved, \edt{which indicates} improved resolution of the local flow physics from the $\text{k–}\omega$ model. However, despite these differences near the blade surface, the downstream vortices generally maintain a consistent shape and trajectory across all grid levels\edt{. It suggests} that the wake features are relatively insensitive to further grid refinement and the accuracy of the $\text{k–}\epsilon$ model. \edt{Furthermore, we also conduct} the time-step independence analysis \edt{using} $t_s = 1800$, $2000$, and $2600$ \edt{time steps per revolution of the propellor}. To further quantify the results of the grid-convergence and timestep-independence studies, the relative error ($\Delta{E}$) between different grids and time step sizes is \edt{defined using the following expression:} 

\begin{eqnarray}
\Delta E = \left|\frac{X_{j}-X_{i}}{X_{j}}\right| \times 100\%
\label{eq:deviation_time}
\end{eqnarray}

\noindent where $j$ denotes the reference case against which case $i$ is compared, and $X$ represents the parameter under consideration. In the present study, case $j$ corresponds to Grid~03, while case $i$ corresponds to Grid~01 and Grid~02. The parameter $X$ represents the mean thrust coefficient ($\overline{C_T}$). The relative errors obtained from both the grid-independence and time-step independence analyses are summarized in Table~\ref{tab:grid_time}. The\edt{se} results indicate that the error between \edt{$\overline{C_T}$ obtained from} Grid~01 and Grid~03 is comparable to that \edt{from} Grid~02 and Grid~03. Similarly, the variation in error across $1800$, $2000$, and $2600$ time-steps per \edt{rotational} cycle remains negligible. Therefore, \edt{we employ Grid~01 for our actual simulations related to this work while using} $1800$ time-steps \edt{in one revolution of the propellor.}

\begin{figure}[h!]
    \centering
    \includegraphics[width=1\linewidth]{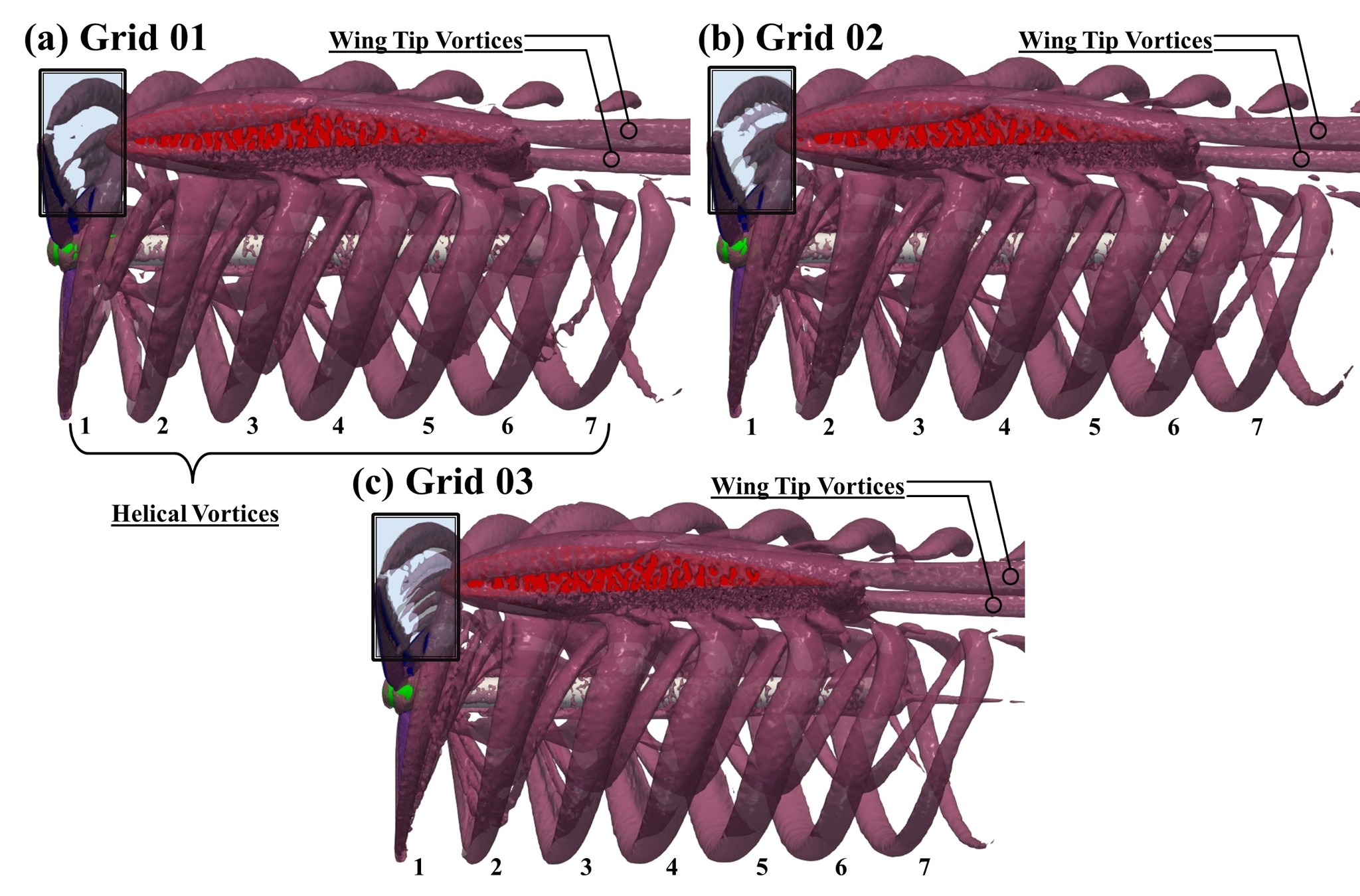}
    \caption{\edt{Comparison of our results obtained from the three grids used for the mesh-convergence tests, where} iso-surfaces of vortical structures \edt{($Q=1000$) are presented} for (a) Grid~$01$, (b) Grid~$02$, and (c) Grid~$03$, highlighting the \edt{development of} wing tip vortices and wake}
    \label{fig:gridInd}
\end{figure}

\begin{table}[h!]
\caption{\label{tab:grid_time}
Comparison of $\Delta E$ for grid-convergence and time-step independence \edt{tests}}
\centering
\begin{tabular}{lcccc}
\hline
 & \multicolumn{2}{c}{Grids (\%)} & \multicolumn{2}{c}{Time-step (\%)} \\
\cline{2-3} \cline{4-5}
Quantity & 01 \& 03 & 02 \& 03 & 1800 \& 2600 & 2000 \& 2600 \\
\hline
$\overline{C_T}$ & 4.01 & 3.40 & 0.10 & 0.03 \\
\hline
\end{tabular}
\end{table}

\begin{figure}[h!]
    \centering
    \includegraphics[width=0.6\linewidth]{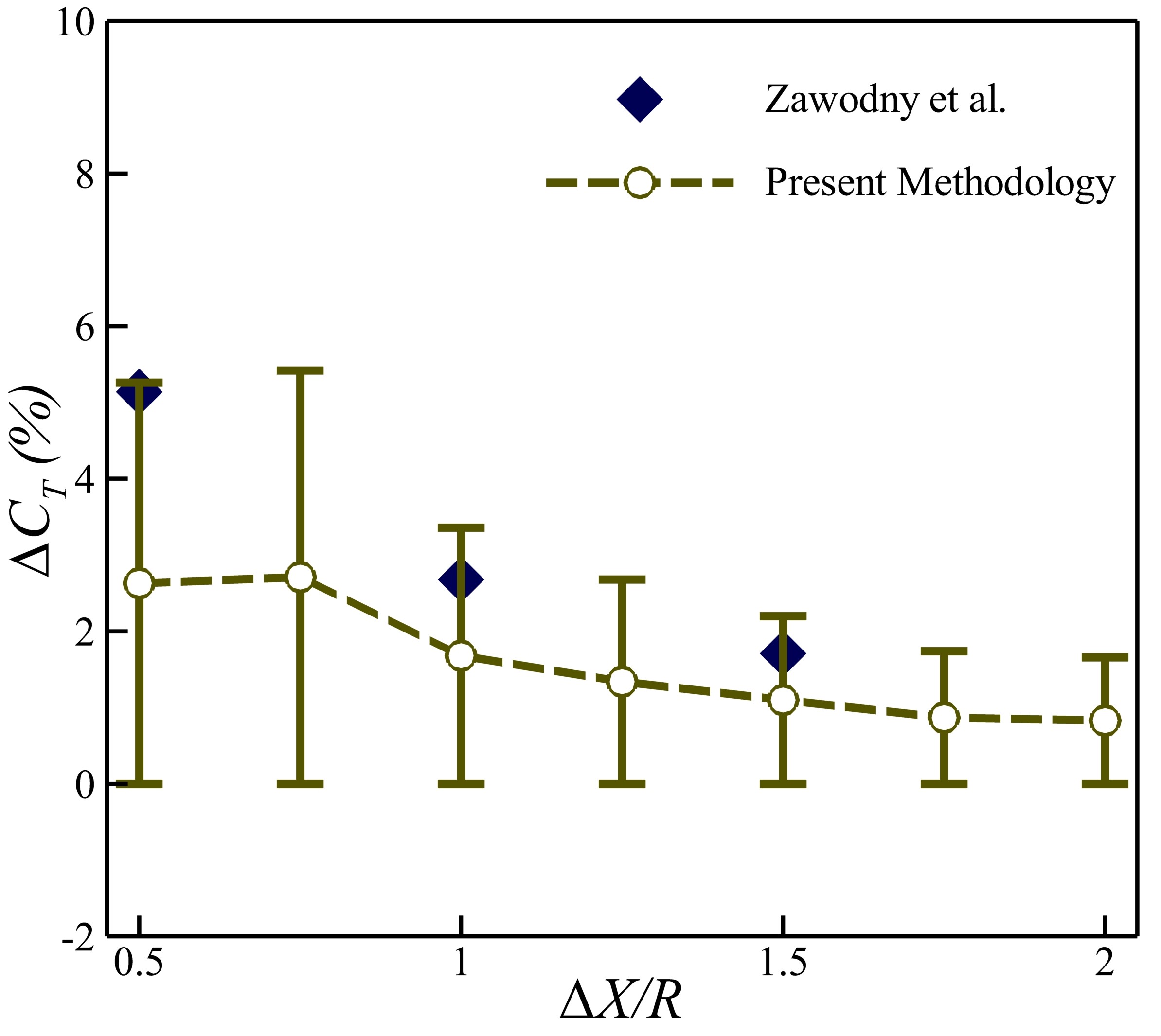}
    \caption{Validation of thrust with varying $\Delta X/R$. Comparison of $\Delta C_T\%$ between the present methodology and Zawodny \textit{et al.}~\cite{zawodny2021aerodynamic}.}
    \label{fig:validation}
\end{figure}

\edt{Next, we validate} robustness \edt{and accuracy} of \edt{our} computational framework \edt{and the used} turbulence model \edt{by comparing our results with those from} Zawodny \textit{et al.} \cite{zawodny2021aerodynamic}. Their work investigates a $\mbox{VTOL}$ configuration with propeller–wing interaction in the near wake, examining the effect of wing\edt{'s location} relative to centerline \edt{of the nacelle} ($\Delta Y/R$) and the \edt{axis of the} propeller ($\Delta X/R$). \edt{Following the experiments from Zawodny \textit{et al.} \cite{zawodny2021aerodynamic}}, \edt{we carry out our simulations using} an advance ratio of $J = 0.592$ with an angular velocity of $\Omega = 6000~\text{RPM}$ \edt{of the propellor} for configurations with and without the wing. \edt{For the cases with the wing, it is placed at different locations ($\Delta X/R = 0.5 - 2.0$). \edt{How the wing's position influences the aerodynamic performance of the propellor} is quantified using the percentage difference in thrust coefficients, $\Delta C_T~\%$, \edt{obtained from} configurations with and without the wing. \edt{Figure}~\ref{fig:validation} presents $\Delta C_T~\%$ as a function of $\Delta X/R$ based on their reported experimental data. The results indicate that increasing $\Delta X/R$ reduces the difference in thrust coefficient, as the influence of the wing on the propeller\edt{'s} wake diminishes. Figure}~\ref{fig:validation} shows that the present methodology captures the overall decreasing trend of $\Delta C_T~\%$ with increasing $\Delta X/R$, demonstrating \edt{a} good agreement \edt{between our results and those from} the experimental data reported by Zawodny \textit{et al.} \cite{zawodny2021aerodynamic}. Minor deviations \edt{might be caused because} they \edt{utilized} a Mejzlik $16\times11$ three-bladed carbon fiber propeller and a $NACA~63_{2}\text{-}215$ MOD B airfoil\edt{-}based wing, which differs from the configuration employed in the present study. Additionally, \edt{radius of the} propeller in their work is reported as $R = 0.2$, whereas \edt{we employ} $R = 0.05$ \edt{in the present work. It is important to mention that} the \edt{aerodynamic performance metrics} can be \edt{compared here based on similarity (affinity) laws}, which corresponds to maintaining the same advance ratio \edt{\citep{Hunsaker1915similitude, vincenti1979air}}. 

\section{RESULTS \& DISCUSSION}

To facilitate navigation through the findings of \edt{our investigations}, \edt{we divide this} section in two subsections. Section~\ref{sec:1} focuses on the analysis and results of the propeller-wing configuration in cruise, whereas Section~\ref{sec:2} focuses on the same configuration in hover. Although the two subsections form a continuous analysis and discussion, this division provides a clear distinction between the two operating conditions.

\begin{figure}[h!]
    \centering
    \includegraphics[width=1\linewidth]{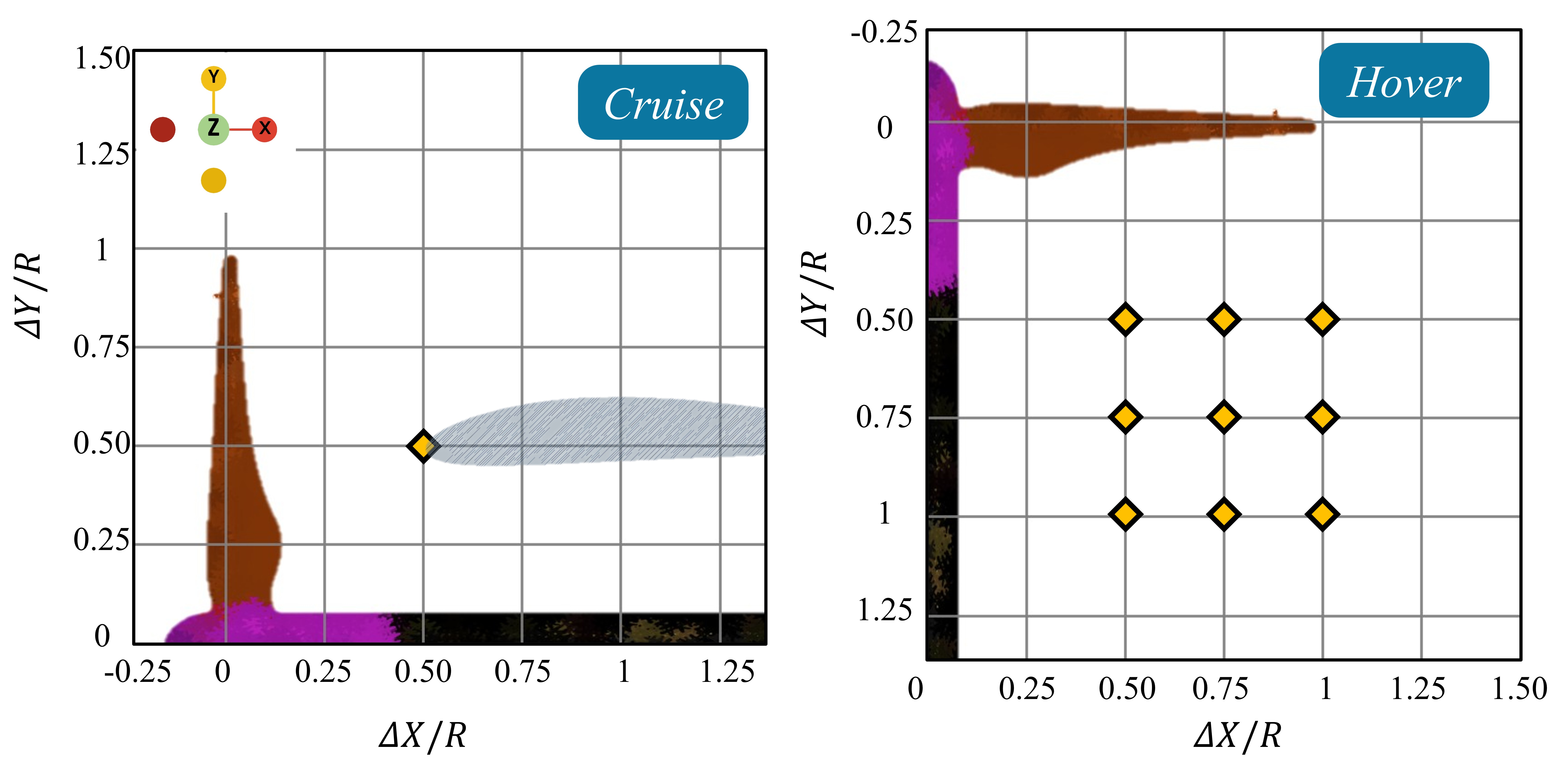}
    \caption{\edt{Placement of the wing} in (a) cruise at fixed $\Delta X/R=\Delta Y/R=0.50$ and (b) hover for varying $\Delta X/R$ and $\Delta Y/R$.}
    \label{fig:wingplacement}
\end{figure}

In \edt{a} VTOL aircraft, the placement of the wing within the near wake of the propeller is important\edt{,} because the wing is the first downstream \edt{structural} component to interact with the propeller slipstream. As discussed earlier, the experimental \edt{work} by Zawodny \textit{et al.} \cite{zawodny2021aerodynamic} quantified the effects of \edt{a} wing\edt{'s} placement at different axial and vertical locations relative to the propeller\edt{'s} centerline and \edt{its} axis of rotation. \edt{It} motivates the present investigation of the propeller-wing interactions during cruise at different freestream velocities and advance ratios. Figure~\ref{fig:wingplacement}a shows \edt{location of} the fixed wing used \edt{employed for our analysis for the cruising condition}. For the \edt{hovering} configuration, the analysis focuses on effects of \edt{the} wing\edt{'s} placement at different positions in the $X$- and $Y$-directions, as shown in Fig.~\ref{fig:wingplacement}b, over a range of rotational speeds \edt{of the propellor}.

\subsection{Propeller-Wing Configuration in Cruise}
\label{sec:1}

For this part of \edt{our work}, the vertical and axial wing positions are fixed at $\Delta Y/R=0.50$ and $\Delta X/R=0.50$, respectively. The freestream velocities are selected as $U_{\infty}~\mathrm{m/s}=4$, $8~\mathrm{m/s}$, and $12~\mathrm{m/s}$. Previous studies reported that increasing the advance ratio from $0.2$ to $0.8$ generally \edt{increased} propulsive efficiency while reducing the thrust coefficient \edt{\citep{abdulrahman2025review, adkins1994design, simmons2021system}}. Accordingly, the advance ratio is varied from $0.2$ to $0.8$ in increments of $0.1$. The operating parameters used in the analysis for cruising are listed in Table~\ref{tab:parameters_cruise}.

\begin{table}[h!]
\caption{\label{tab:parameters_cruise}
Governing parameters for wing-rotor interactions in cruise}
\centering
\begin{tabular}{lcc}
\hline
Parameter & Symbol & Value \\
\hline
Freestream velocity ($m/s$)     & $U_{\infty}$ & $4.0$, $8.0$, $12.0$ \\
Advance ratio            & $J$          & $0.2$--$0.8$ \\
Vertical wing position   & $\Delta X/R$ & $0.50$ \\
Axial wing position      & $\Delta Y/R$ & $0.50$ \\
\hline
\end{tabular}
\end{table}

Smaller $UAVs$ and drones experience lower Reynolds number compared to their larger counterparts. The propellers with larger blade diameter often experience Reynolds number of over \edt{$10^8$} operating at $\Omega<6000$ \cite{deters2014reynolds, casalino2021definition}. Based on the relationship between the angular velocity, freestream velocity and the advance ratio, the calculated $\mbox{RPM}$ alongside their corresponding Reynolds number is shown in Fig.~\ref{fig:framework}, which shows the operating $\Omega$ of the propeller, \edt{ranging from $5000$ to $36,000$}, based on the advance ratio required to maintain $\mbox{Re}$ less than \edt{$10^8$}.

\begin{figure}[h!]
    \centering
    \includegraphics[width=1\linewidth]{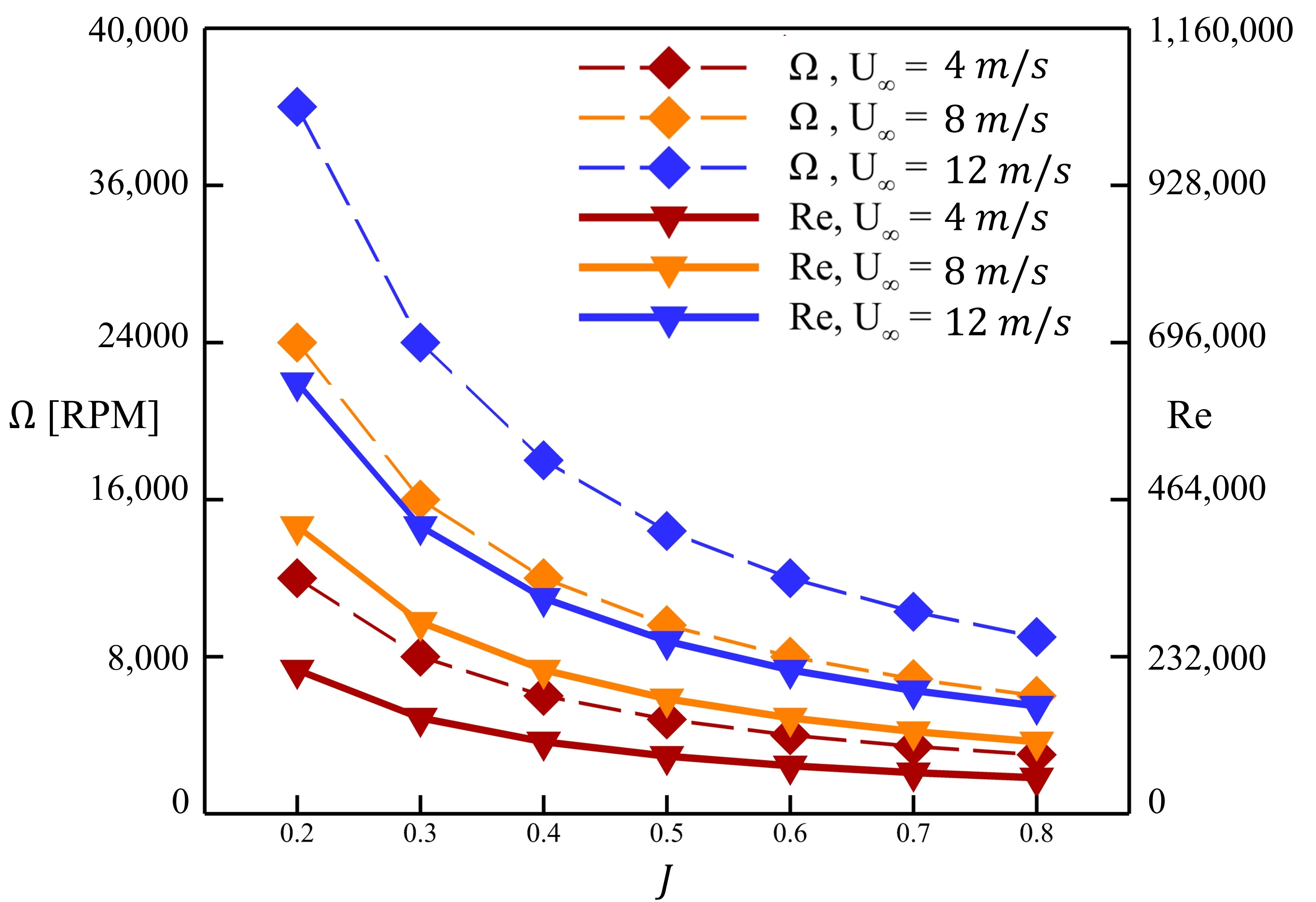}
    \caption{Variation of rotational speed $\Omega$ and Reynolds number $\mbox{Re}$ \edt{versus} $J$ for different freestream velocities}
    \label{fig:framework}
\end{figure}

The efficiency of the propeller and the thrust coefficient versus $J$ are plotted in Fig.~\ref{fig:eff}. \edt{These plots clearly exhibit} the region of maximum efficiency and the corresponding thrust coefficient. The trend observed \edt{here} follows the trend of optimal efficiency close to the advance ratio around $J\approx0.7$, and \edt{these} trends \edt{also} show that the efficiency starts \edt{dropping} with further increasing $J$ to $0.8$.  

\begin{figure}[h!]
    \centering
    \includegraphics[width=1\linewidth]{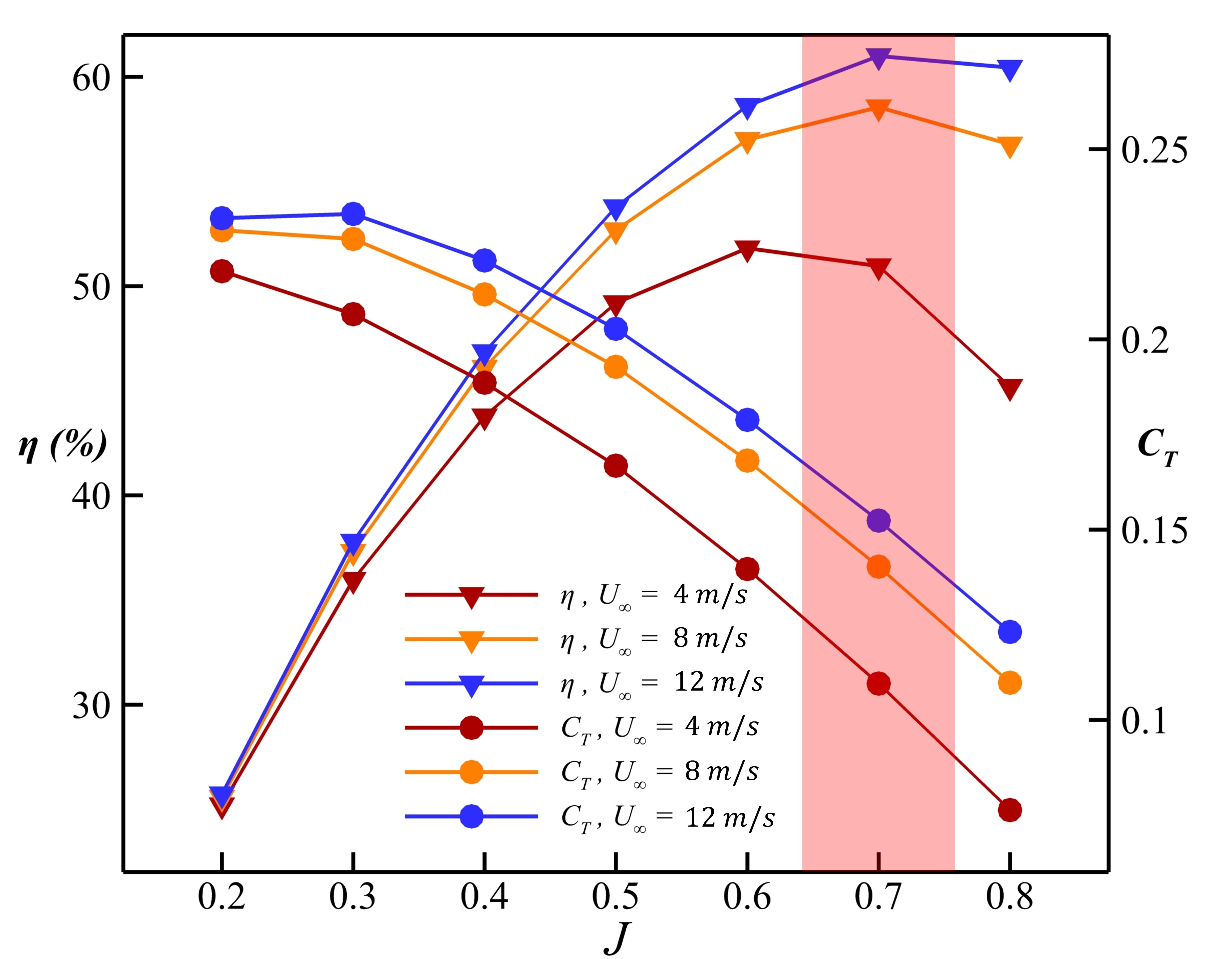}
    \caption{Variation of propulsive efficiency ($\eta$) and thrust coefficient ($C_T$) with $J$ for different freestream velocities\edt{, where} the shaded region indicates \edt{values of $J$ to attain} peak efficiency}
    \label{fig:eff}
\end{figure}

\begin{figure}[h!]
    \centering
    \includegraphics[width=1\linewidth]{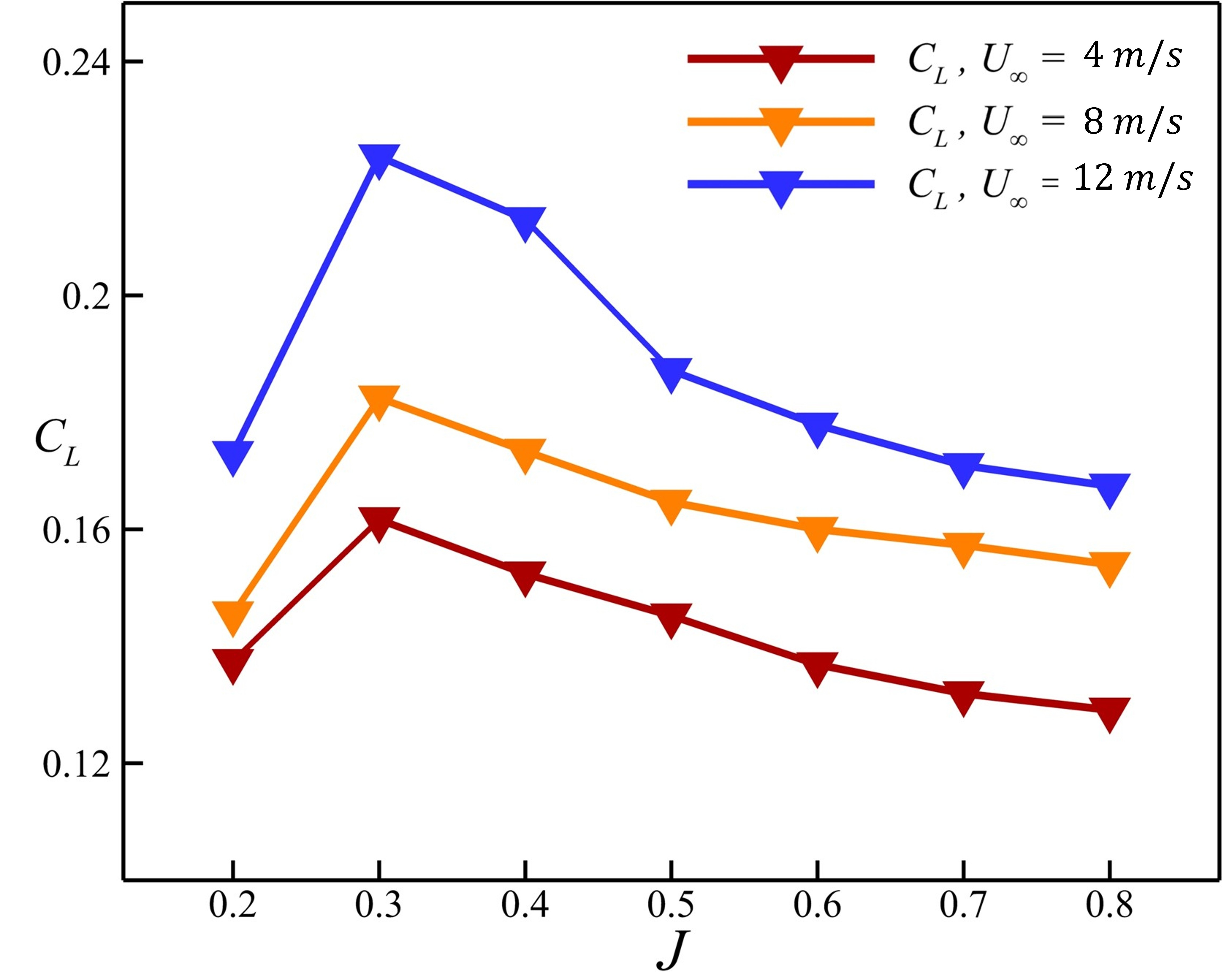}
    \caption{Variation of lift coefficient $C_L$ with advance ratio $J$ for different freestream velocities.}
    \label{fig:liftCoefficient}
\end{figure}

Under the operating condition of $J=0.7$, the present study reports operating efficiencies of $50\%$, $58\%$, and $61\%$ for the considered freestream velocities, corresponding to \edt{$\Omega$ of $3429~\mathrm{RPM}$, $6857~\mathrm{RPM}$, and $10286~\mathrm{RPM}$}, respectively. These results also indicate that \edt{propellers of drones} at this scale may operate at relatively high rotational speeds in the proposed $VTOL$ configuration. Furthermore, the cambered wing considered \edt{here} generates positive lift across all investigated advance ratios. Figure~\ref{fig:liftCoefficient} presents the time-averaged lift coefficient ($C_L$) of the wing at different values of $J$ and $U_{\infty}$. The lift coefficient exhibits a decreasing trend similar to that of the thrust coefficient shown in Fig.~\ref{fig:eff}. However, the operating condition corresponding to the maximum propulsive efficiency does not necessarily produce the \edt{maximum lift by the wing}. As discussed later, operation at higher advance ratios produces a narrower and more coherent jet-like propeller slipstream that retains its streamwise momentum farther downstream after impinging on the wing. The resulting localized flow acceleration beneath the wing reducing the pressure on the bottom-side surface, thereby decreasing the pressure difference across the wing and producing a corresponding reduction in the net $C_L$.

At a given $U_{\infty}$ and \edt{diameter of the propellor}, lower $J$ correspond to higher $\Omega$. \edt{A} propeller operating at low $J$ generally associates with reduced \edt{$\eta$} \edt{due to} increased blade loading, induced velocity, and energy losses within the slipstream. \edt{Now, we link link the aerodynamic performance metrics with unsteady flow dynamics around the propellor and the wing}. \edt{Figures~\ref{fig:avg}a and \ref{fig:avg}c \edt{present} the propeller slipstream visualized using a $Q$-criterion isosurface of $Q=30,000$ at a cruise velocity of $U_{\infty}=12~\mathrm{m/s}$ with $J=0.3$ and $J=0.7$, respectively. Figures~\ref{fig:avg}b and \ref{fig:avg}d show the corresponding contours of the cycle-averaged streamwise velocity. At $J=0.3$, the slipstream exhibits weak wake coherence, as shown in Fig.~\ref{fig:avg}a. Successive tip vortices interact through an early-stage leapfrogging mechanism, highlighted by the yellow box. \edt{Because of the} high \edt{$\Omega$ here}, successive tip vortices \edt{from blades of the propellor} are generated at a rate greater than their downstream convection, resulting in a reduced helical pitch and closely spaced vortex filaments \edt{\citep{gudmundsson2013general}}. The reduced spacing promotes strong mutual induction between successive tip vortices, leading to vortex pairing, leapfrogging, and roll-up \edt{\citep{felli2011mechanisms, lugt1996introduction}}. These interactions may subsequently produce vortex merging, instability, and accelerated wake breakdown, thereby increasing turbulent mixing and dissipating the coherent kinetic energy of the propeller\edt{s} slipstream \edt{\citep{yamada1978preliminary, felli2011mechanisms, baek2015effects}}. The resulting increase in induced and mixing losses contributes to the lower propulsive efficiency observed at low advance ratios in Fig.~\ref{fig:eff}. \edt{Moreover, we} \edt{also identify} in the \edt{contour plot of} cycle-averaged velocity in Fig.~\ref{fig:avg}b that the propeller produces a relatively \edt{diffused} wake with weak jet-like axial momentum \edt{under these flow and kinematic conditions}. Contrarily, the wake retains stronger coherence over a greater downstream distance at $J=0.7$, as shown in Fig.~\ref{fig:avg}c, thereby reducing vortex-vortex \edt{interactions} and associated energy losses. \edt{A contour plot of} The cycle-averaged velocity in Fig.~\ref{fig:avg}d consequently exhibits a narrower and more concentrated jet-like streamwise velocity distribution. Following interactions with the wing, the slipstream retains substantial axial momentum along both the pressure and suction sides of the wing. At intermediate advance ratios ($J=0.6 - 0.8$), the increased axial convection produces a larger helical pitch and greater spacing between successive tip vortices, reducing vortex-vortex interactions and allowing the slipstream to retain its coherent jet-like structure over a longer distance downstream. This condition promotes more effective streamwise momentum transfer with lower induced and rotational losses \edt{in the flow}, resulting in higher \edt{$\eta$}. \edt{However,} the relative blade loading and thrust coefficient decrease substantially \edt{for excessively high advance ratios ($J>0.8$)} as the inflow velocity becomes large relative to the blade\edt{'s} rotational speed. Consequently, both the thrust generated by the propeller and its propulsive efficiency decrease, \edt{as demonstrated by the data presented} in Fig.~\ref{fig:eff}.

}

\begin{figure}[h!]
    \centering    
    \includegraphics[width=1\linewidth]{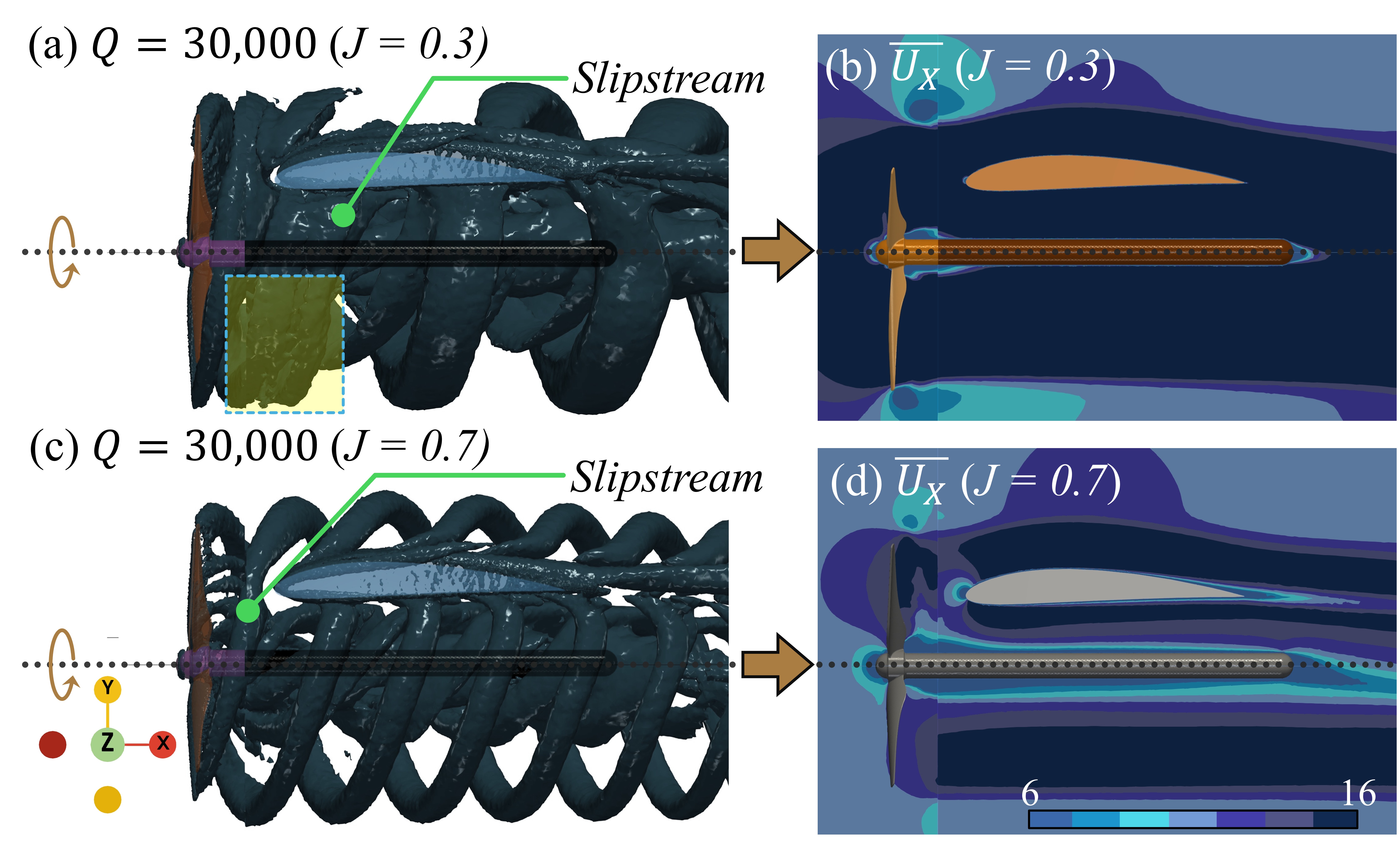}
    \caption{Sectional view of Propeller-wing configuration at $Q=30,000$, and cycle-averaged velocity contours in streamwise direction for two operating conditions: (a, b) $J=0.3$ and (c, d) $J=0.7$, at $U_\infty=12 ~m/s$, respectively.}
    \label{fig:avg}
\end{figure}

From Fig.~\ref{fig:eff}, the peak propulsive efficiency generally occurs at $J=0.7$. For $U_{\infty}=4~\mathrm{m/s}$, the maximum efficiency is observed at $J=0.6$\edt{. Nevertheless, we examine} the cases at $U_{\infty}=4~\edt{\mathrm{m/s}}$, $8~\edt{\mathrm{m/s}}$, and $12~\mathrm{m/s}$ at $J=0.7$ \edt{for consistency in the comparative analysis}. The propeller slipstreams \edt{are} visualized using a $Q=30,000$ isosurface in Figs.~\ref{fig:comparison}a, \edt{~\ref{fig:comparison}b, and} \ref{fig:comparison}c for $U_{\infty}=4~\edt{\mathrm{m/s}}$, $8~\edt{\mathrm{m/s}}$, and $12~\mathrm{m/s}$, respectively. The isometric views show that, at a constant $J$, increasing the cruise velocity produces stronger vortical structures within the slipstream. Since maintaining a constant $J$ at a higher $U_{\infty}$ requires a proportional increase \edt{in $\Omega$ of the} propeller, \edt{velocity of the blade's tip} and kinetic energy \edt{of the wake} also increase \edt{\citep{baek2015effects, adkins1994design, felli2011propeller, felli2021underlying}}. This trend demonstrates that operation near the optimum \edt{$J$} \edt{maintains} a coherent slipstream over a range of cruise velocities. As shown in Fig.~\ref{fig:eff}, the configuration operating at $U_{\infty}=12~m/s$ and $J=0.7$ produces a thrust coefficient comparable to that obtained at $U_{\infty}=4~m/s$ and $J=0.5$, while achieving substantially higher \edt{$\eta$}.

\begin{figure}[h!]
    \centering    \includegraphics[width=1\linewidth]{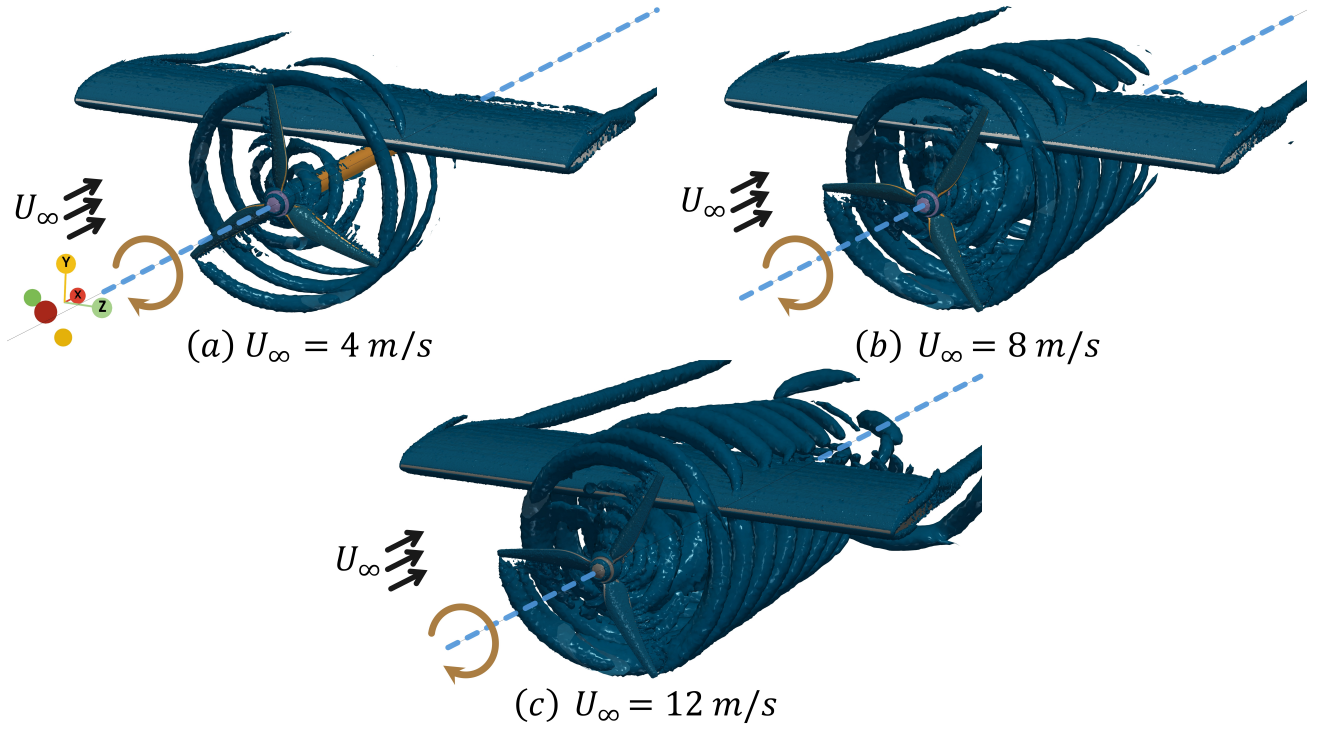}
    \caption{Iso-surfaces of propeller slipstream at $Q=30,000$, and $J=0.7$ for cruising velocities $U_\infty$ (a) $4~m/s$ , (b) $8~m/s$, and (c) $12~m/s$.}
    \label{fig:comparison}
\end{figure}

\edt{Next, we focus on analyzing} interaction\edt{s} between the propeller slipstream and the downstream wing in the VTOL configuration\edt{during cruise}. A propeller typically generates a helical wake that convects downstream while retaining its coherence until it encounters any object far wake or gradually diffuses through viscous and turbulent mixing. When a wing is \edt{located} within the near wake, the slipstream is strongly perturbed \edt{that alters} the trajectory, coherence, and evolution of the vortices. Muscari \textit{et al.} \cite{muscari2017analysis}, Felli \textit{et al.} \cite{felli2011propeller, felli2009experimental}, and Liu \textit{et al}, \cite{liu2004blade} investigated similar interactions between \edt{the wake of a marine propellor} and downstream rudders. Although the present study considers a propeller-wing configuration for aerial vehicles, the underlying mechanisms governing vortex-wing interaction remain closely related to those \edt{observed} for propeller-rudder interactions.

\begin{figure}[h!]
    \centering
    \includegraphics[width=1\linewidth]{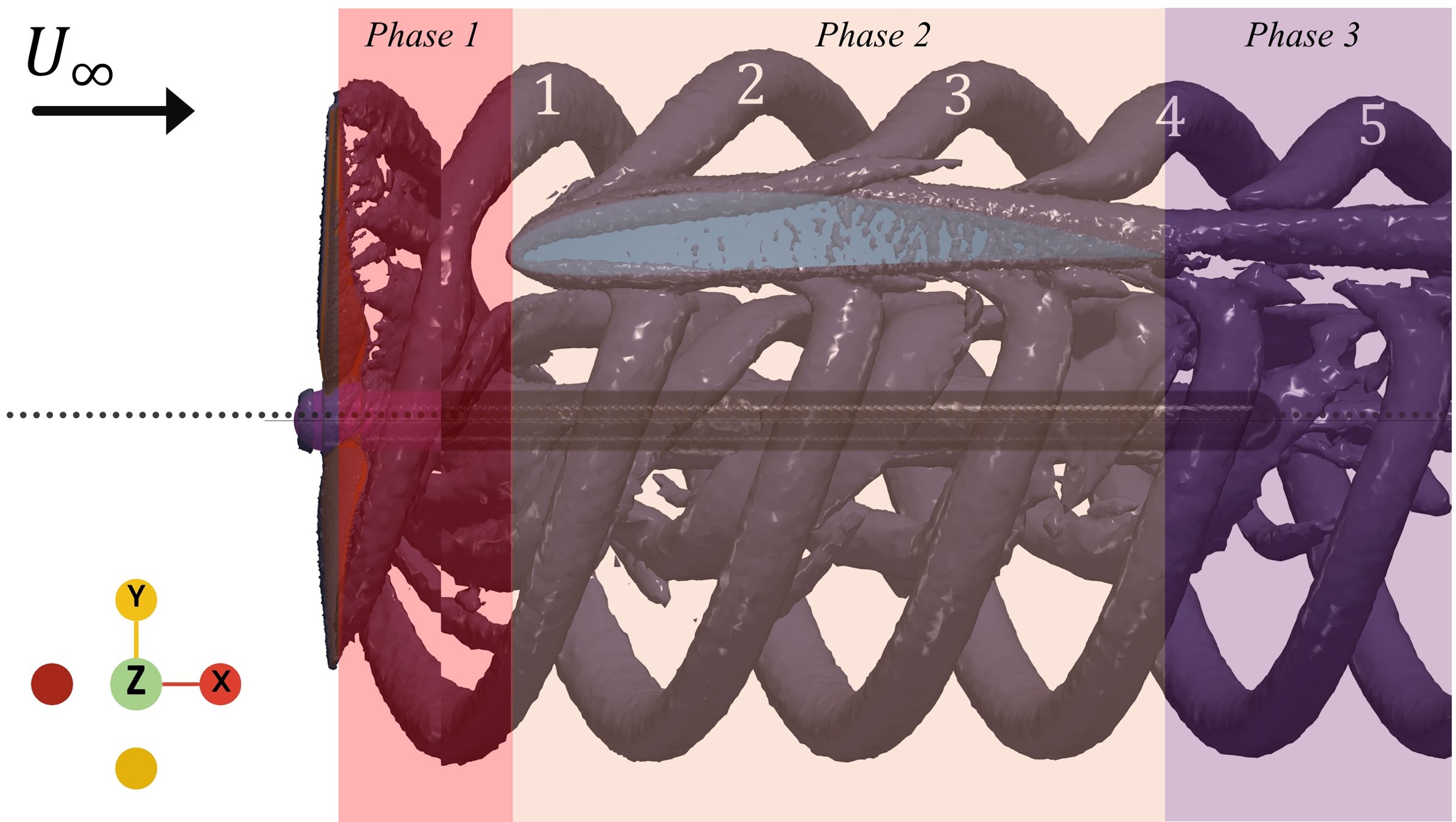}
    \caption{Side view of the tip vortex evolution ($Q=30,000$) and its interaction with the wing, illustrating approaching phase (Phase~1), interaction phase (Phase~2), and the convection phase (Phase~3).}
    \label{fig:wingLiftside}
\end{figure}

Following Muscari \textit{eta al.} \cite{muscari2017analysis}, the \edt{interactive region} between the propeller slipstream and the wing is divided into three phases. \edt{These} are illustrated using \edt{vortex visualizations in} the side-view \edt{of the whole aerodynamic structure} \edt{using isosurfaces of} $Q=30,000$ in Fig.~\ref{fig:wingLiftside} for the cruise \edt{condition} at $U_{\infty}=12~\mathrm{m/s}$ and $J=0.7$. This case is chosen for two reasons. \edt{Firstly}, the configuration operating at $U_{\infty}=12~\mathrm{m/s}$ exhibits the highest propulsive efficiency among all cases considered in \edt{our current work}. \edt{Secondly, wake of the propellor} at the peak-efficiency conditions for all cruise velocities exhibit similar slipstream, with the primary difference being the strength of the downstream vortical structures. \edt{In Fig.~\ref{fig:wingLiftside},} phases~$1$, $2$, and $3$ correspond to the approaching phase (highlighted in red), the interaction phase (highlighted in beige), and the convection phase (highlighted in \edt{purple}), respectively. \edt{It is evident that} the propeller\edt{'s} wake retains characteristics similar to those of an isolated propeller wake~\cite{magionesi2018modal, yu2026symmetric}, despite the presence of a wing within the near-wake downstream. 

\begin{figure}[h!]
    \centering
    \includegraphics[width=1\linewidth]{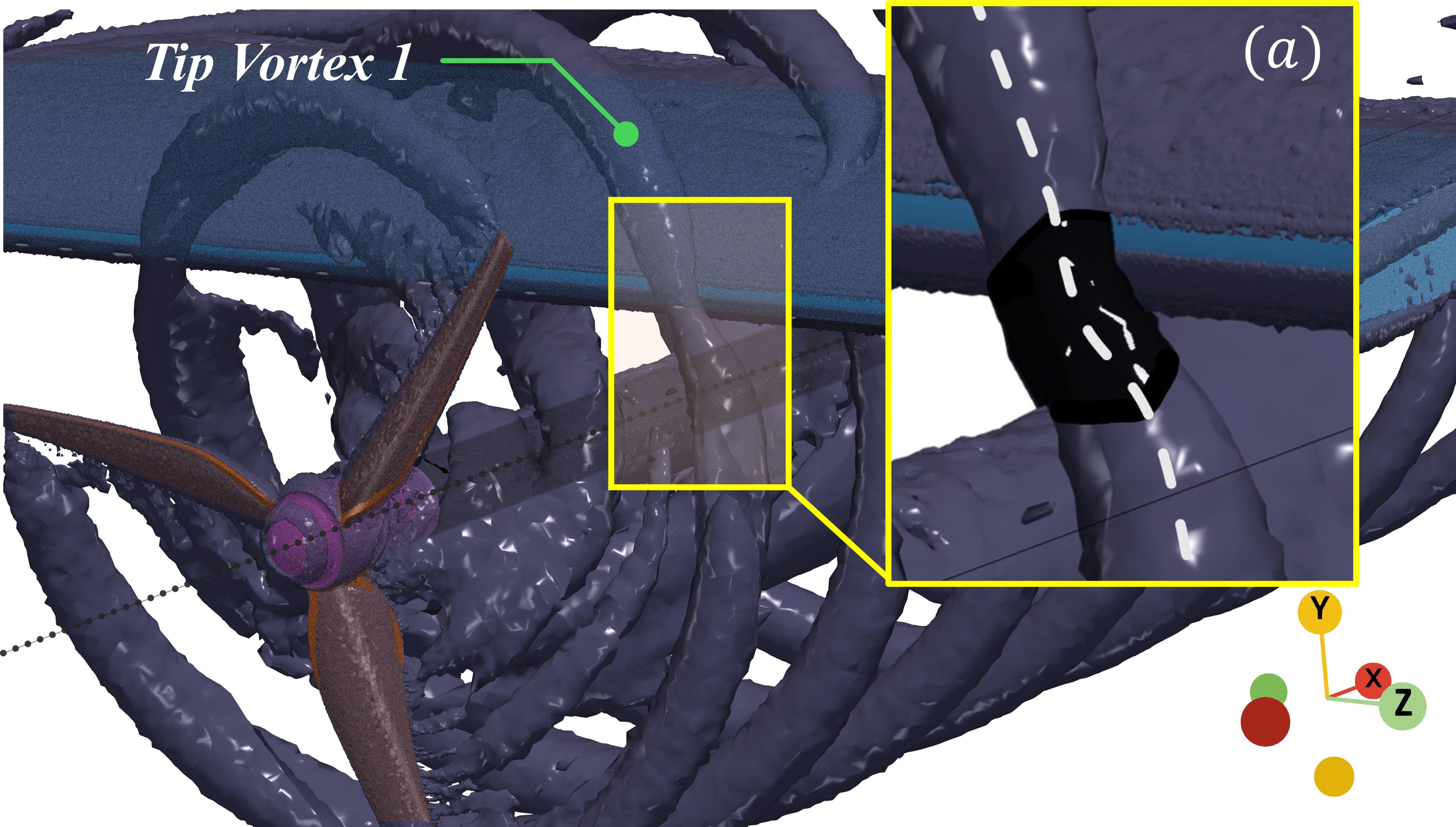}
    \caption{\edt{A zoomed-in view} of vortex–wing interaction highlighting localized vortex deformation and breakdown near the wing ($Q=30,000$)}
    \label{fig:iso}
\end{figure}

During the approaching phase (Phase~$1$), \edt{Muscari \textit{eta al.} \cite{muscari2017analysis}} observed \edt{perturbations} in the tip-vortex at \edt{a} distance of up to $0.3R$ upstream of the wing\edt{'s} leading edge. In comparison, the propeller\edt{'s} rotational speed is considerably higher in the present study, and the perturbation \edt{in the wake} occurs immediately before flow impingement. Figure~\ref{fig:iso} shows an isometric view of the propeller \edt{and its} wake and captures the tip vortex, labelled \edt{as} ``Tip Vortex~1'', immediately before its interaction with the wing. The enlarged view in Fig.~\ref{fig:iso}a clearly shows the stretching of the vortex as it approaches the wing\edt{'s} leading edge. 


\begin{figure}[h!]
    \centering
    \includegraphics[width=1\linewidth]{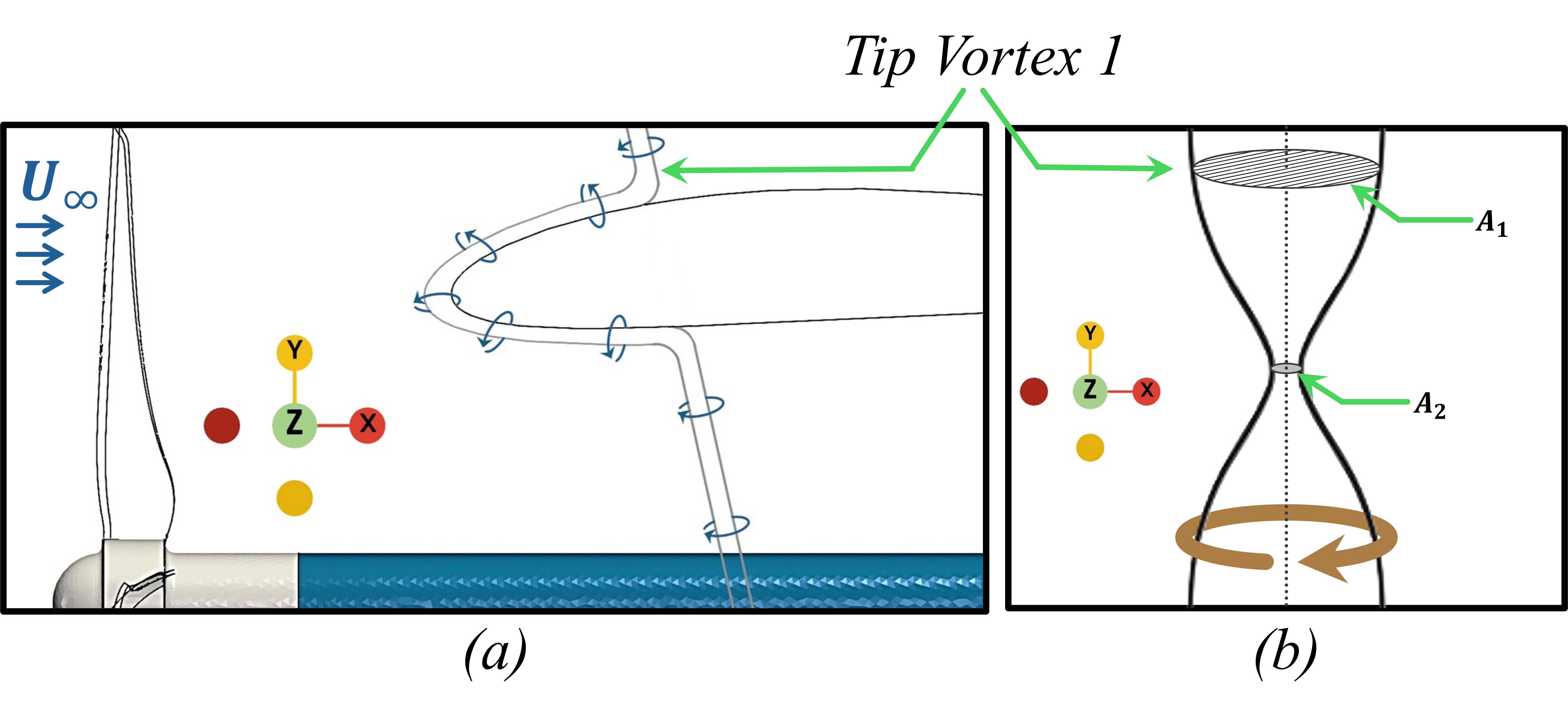}
    \caption{Tip-vortex schematic showing (a) the sectional view during the approaching phase and (b) vortex stretching.}
    \label{fig:schematic2}
\end{figure}

\edt{We further illustrate this vortex activity schematically in Fig.}~\ref{fig:schematic2} \edt{that shows} evolution of the vortex shed from the blade \edt{of the propellor} and its interaction with the wing during \edt{its} impingement. In Fig.~\ref{fig:schematic2}a, Tip Vortex~1 from Fig.~\ref{fig:iso} is isolated and presented from the side view. \edt{This schematic} shows the stretched vortex structure resulting from its interaction with the wing. The arrows surrounding the vortex indicate the direction of rotation about the vortex filament. Upon impingement, the vortex does not immediately dissipate but undergoes stretching. \edt{This observation is} consistent with \edt{those from Muscari \textit{et al.} \cite{muscari2017analysis}}. As illustrated in Fig.~\ref{fig:schematic2}b, axial stretching of the vortex near the wing leading edge reduces its core cross-sectional area ($A_2$) and increases its vorticity \edt{\citep{green2012fluid}}. Such vortex stretching \edt{was} widely reported in the literature \edt{\citep{felli2011propeller, muscari2017analysis, marshall1996penetration}} and is fundamentally consistent with Helmholtz's vortex laws \edt{\citep{von1867handbuch}}.

\begin{figure}[h!]
    \centering
    \includegraphics[width=1\linewidth]{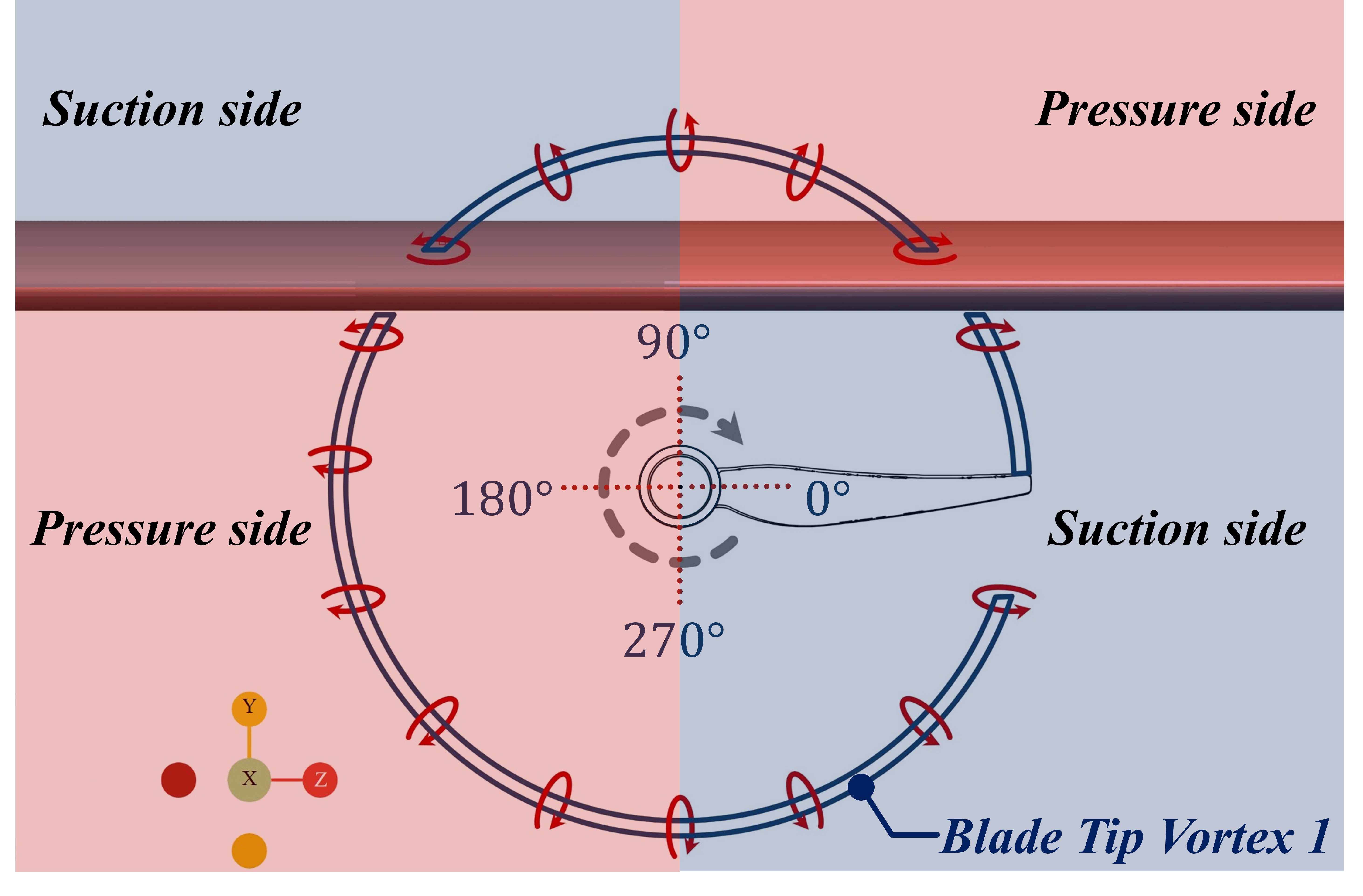}
    \caption{Front-view schematic of the \edt{propellor} with a single blade at various angular positions, showing the vortex structure and the corresponding pressure and suction sides of the wing}
    \label{fig:InteractionPhaseSchematic}
\end{figure}

The interaction phase (Phase~$2$ \edt{in Fig.~\ref{fig:wingLiftside}}) begins when the stretched tip vortex encounters the wing\edt{'s} boundary layer and impinges on the leading edge. The vortex subsequently bifurcates into two branches that convect downstream. Although local interactions between the tip vortex and the wing boundary layer may influence the near-wall flow, their contribution is considered secondary to the large-scale deformation and splitting of the vortex and is\edt{,} therefore\edt{,} not examined separately in the present study. Figure~\ref{fig:InteractionPhaseSchematic} presents a front-view schematic of the propeller-wing configuration. To illustrate the interaction between the tip vortex and the wing, only one propeller blade is shown at an angular position of $0^\circ$. The schematic also identifies the pressure distribution associated with the vortex-wing interaction. As the propeller rotates and the slipstream impinges on the wing, regions of relatively high and low pressure develop over the wing surfaces. For consistency in the terminology used throughout \edt{this manuscript}, the pressure side is shown in red, whereas the low-pressure, or suction, side is shown in blue. The rotational direction of the tip vortex immediately before impingement is also illustrated in the schematic shown in Fig.~\ref{fig:InteractionPhaseSchematic}. During \edt{this} phase, \edt{branches of} the bifurcated tip-vortex convect downstream while largely preserving symmetry about the propeller\edt{'s} axis of rotation. An exception is the branch passing over the suction side of the wing, which exhibits pronounced asymmetric displacement. The symmetry breaking and subsequent trajectory of the bifurcated vortices are primarily governed by three mechanisms. \edt{Muscari \textit{et al.} \cite{muscari2017analysis}, Green \textit{et al.} \cite{green2012fluid}, and Felli \textit{et al.} \cite{felli2009experimental, felli2011mechanisms}} attributed this behavior to the ``vortex-image effect'' and the ``blockage effect''. \edt{In our current work}, a third mechanism associated with streamwise momentum convection is introduced to describe the downstream transport of the bifurcated vortex branches. The vortex-image effect arises when a vortex approaches an impermeable surface and experiences an induced velocity equivalent to that produced by an image vortex of equal magnitude and opposite circulation located on the opposite side of the surface. For the \edt{presently employed} configuration, this interaction induces a displacement of the vortex in the $+Z$ direction over the wing. \edt{On the other hand,} the blockage effect arises because the wing acts as a solid obstruction within the propeller\edt{'s} slipstream. The cross-sectional area of the vortex core interacting with the wing is greatest near the leading edge due to its thickness and progressively decreases as the bifurcated vortex branches convect toward the trailing edge where \edt{thickness of} the wing is minimal. \edt{Equivalently, it can be stated that} the influence of blockage weakens from the leading edge to the trailing edge. This effect promotes the global displacement of the vortex branches on both sides of the wing away from the propeller centerline. The \edt{newly introduced} third mechanism is associated with the axial convection imposed by high-momentum slipstream \edt{from the propellor. Here,} its high rotational speed generates a strong jet-like axial velocity that transports the bifurcated vortex structures downstream. \edt{Hence,} this convective effect drives the vortex branches predominantly in the streamwise $+X$ direction. 

\begin{figure}[h!]
    \centering
    \includegraphics[width=1\linewidth]{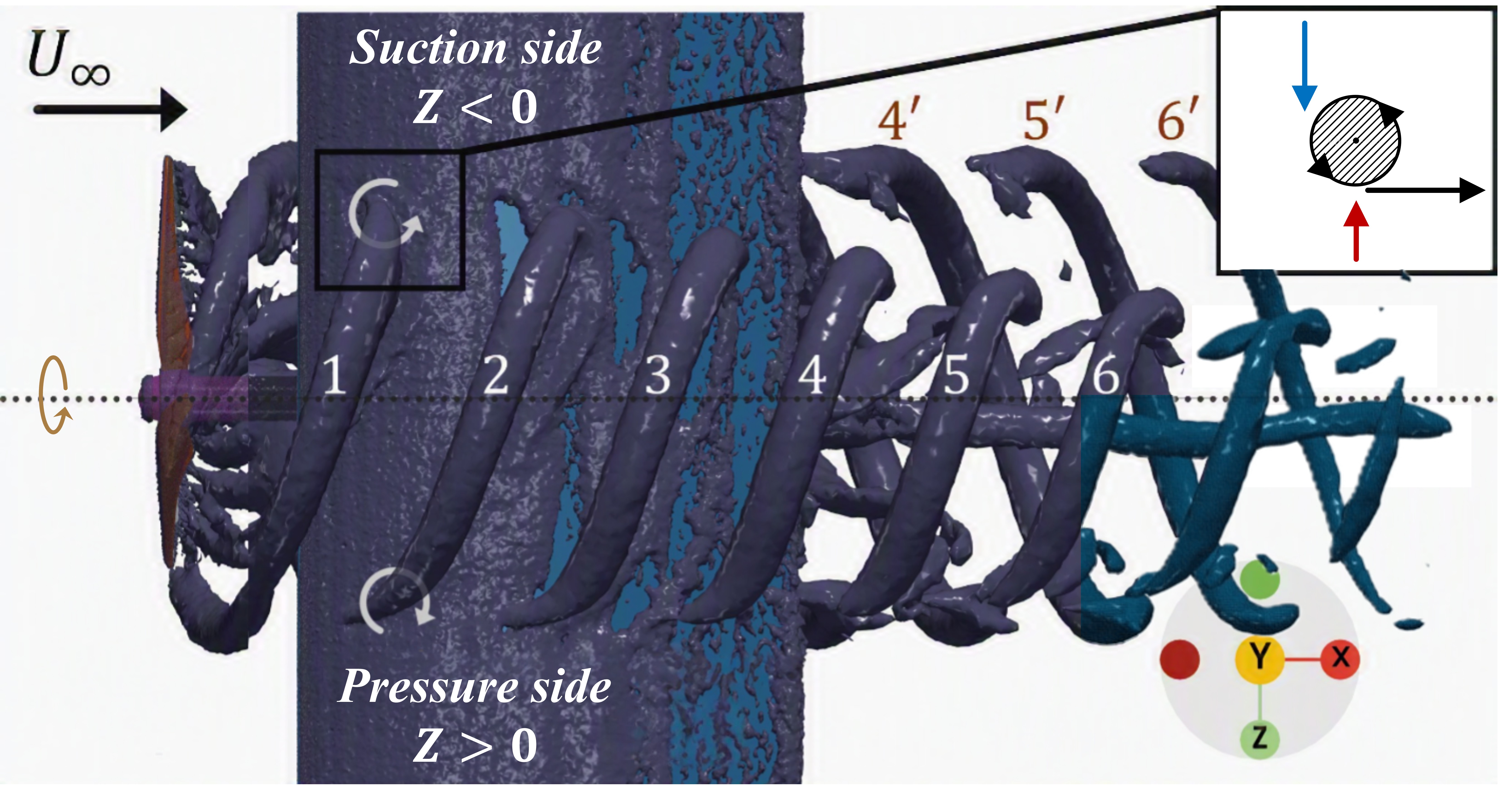}
    \caption{Interaction of the tip vortex (\edt{visualized by isosurfaces of} $Q=30,000$) from the blade of the propeller with the wing, illustrating \edt{distortion in the} vortex and asymmetry across the span ($Z<0$ and $Z>0$)}
    \label{fig:wingLift}
\end{figure}

\edt{To visualize this scenario,} Fig.~\ref{fig:wingLift} shows the \edt{top view of the} propeller-wing configuration. Over the suction side of the wing, \edt{the branches of} the vortex exhibit pronounced asymmetric convection, whereas along the pressure side, the vortices undergo significant initial deflection upon interaction with the wing before convecting approximately straight downstream. The vortices over \edt{the} wing are labeled $1$-$6$, while their corresponding counterparts at the bottom of the wing are labeled $1'$-$6'$. \edt{Please note that the vortices identified as $1'$, $2'$, and $3'$ are under the wing and remain hidden in this top view of the configuration}. The inset in the upper-right corner of Fig.~\ref{fig:wingLift} presents a sectional view schematic of the vortex core and indicates the streamwise momentum convection, induced velocity, and blockage effect using black, blue, and red arrows, respectively. Relative \edt{\dn{approximate} lengths of the arrows} represent the comparative influence of these three mechanisms on the vortex trajectory. On the suction side of the wing, the blockage effect acts away from the propeller\edt{'s} centerline, whereas the image-induced velocity acts toward the centerline. \edt{Therefore,} these two effects oppose each other. Based on the mechanisms established above, vortex~$1$ on the suction side is expected to experience the smallest transverse deflection from its streamwise trajectory. \edt{However}, a finite displacement is still observed. As the blockage effect weakens toward the trailing edge, the image-induced velocity becomes increasingly dominant, causing vortices~$2$-$4$ on the suction side to shift toward \edt{the} centerline \edt{of the propellor}. \edt{Oppositely}, both the blockage effect and the image-induced velocity act away from the propeller centerline on the pressure side. Consequently, vortex~$1$ experiences the largest transverse deflection. As \edt{branches of} the vortex convect toward the trailing edge and the blockage effect diminishes, the magnitude of the deflection gradually decreases, and the vortices recover a predominantly streamwise trajectory. Vortex~$2$ exhibits a pronounced local curvature in its filament near the wing\edt{'s} surface. \edt{It may be a} result from the combined influence of the image-induced velocity and blockage effect. As the blockage effect weakens toward the trailing edge, this curvature gradually relaxes.
 
It is important to note \edt{that} the placement of the wing above the nacelle introduces a vertical asymmetry from the center of the $Y$-$Z$ plane, resulting in a larger vortical structure at the bottom of the wing. In this region of the wake, streamwise momentum convection is highly dominant, as vortices~$1'$-$6'$ retain an approximately symmetric downstream trajectory with limited influence from either the blockage effect or the image-induced velocity. The vortex-image effect is primarily evident during the initial interaction near the wing\edt{'s} leading edge, where vortex~$1'$ exhibits a slight displacement away from the propeller\edt{'s} centerline. \edt{Next, when} the vortex detaches from the wing, it enters the convection phase. The approaching and interaction phases establish the influence of the downstream wing on the vortex trajectory and deformation. During the convection phase, vortices~$1$-$6$ shed from above the wing \edt{exhibit} a strongly asymmetric downstream trajectory\edt{, as presented in Fig.~\ref{fig:wingLift}}. The branches originating from the suction side are displaced toward the propeller centerline, whereas those originating from the pressure side convect predominantly in the $+X$-direction. In contrast, vortices~$1'$-$6'$ from under the wing, on both the suction and pressure sides, retain an approximately streamwise trajectory. Consequently, reconnection between vortices~$1$-$6$ and $1'$-$6'$ is not clearly observed on the $Z<0$ side of the wing, whereas it is evident on the $Z>0$ side through a ``split-branches-rejoin'' like mechanism similar to that reported by Muscari \textit{et al.} \cite{muscari2017analysis}.

An additional mutual-induction effect is observed during the convection phase between the vortices shed above and below the wing on the $Z<0$ side. In Fig.~\ref{fig:wingLift}, the vortex pairs $4$-$4'$, $5$-$5'$, and $6$-$6'$ rotate counter-clockwise. If these paired vortices convected with mirrored symmetry, their mutual induction could promote constructive interaction downstream and \edt{help} sustain higher axial momentum transfer. However, because the upper and lower vortex branches experience different transverse deflections, the induced velocities associated with the pairs $4$-$4'$, $5$-$5'$, and $6$-$6'$ do not remain symmetrically aligned. \edt{Hence,} their interaction amplifies the transverse displacement of the vortices shed over the suction side, as observed in Fig.~\ref{fig:wingLift}. Although \edt{our} present analysis \edt{for the cruise condition} does not consider variations in placement \edt{of the wing with respect the propellor}, Zawodny \textit{et al.} \cite{zawodny2021aerodynamic} reported that positioning the wing closer to the nacelle, corresponding to a smaller $\Delta Y$, produced a higher thrust coefficient than \edt{the} configurations with the wing positioned farther from the nacelle.

\subsection{Propeller-Wing Configuration in Hover}
\label{sec:2}

\edt{The discussion on aerodynamic performance of a propellor-wing arrangement and vortex dynamics around it provides insights for} the influence of advance ratio and freestream velocity on the evolution of the propeller slipstream and its interaction with the downstream wing. \edt{However,} the absence of freestream \edt{in hovering conditions} changes the wake \edt{dynamics} and loading characteristics, \edt{which makes the} propeller\edt{'s} rotational speed and \edt{a} wing\edt{'s} placement the primary governing parameters. \edt{To examine such situations,} the \edt{wing's} vertical and axial positions are varied as $\Delta Y/R=\Delta X/R=0.50$, $0.75$, and $1.00$, respectively. Under this condition, the advance ratio remains zero and cannot be used to vary the operating condition. Therefore, \edt{we vary} the propeller\edt{'s} rotational speed from $8,900\edt{~\mbox{RPM}}$ to $24,200~\mbox{RPM}$. The operating parameters used for this part of \edt{our work} are listed in Table~\ref{tab:parameters_hover}.

\begin{table}[h!]
\caption{\label{tab:parameters_hover}
Governing parameters for wing-rotor interactions in hover.}
\centering
\begin{tabular}{lcc}
\hline
Freestream velocity ($m/s$)     & $U_{\infty}$ & $0.0$ \\
Advance ratio            & $J$          & $-$ \\
Propeller Rotational Speed (RPM)           & $\Omega$          & $8,900$; $11,430$; $15,740$; $24,200$ \\
Vertical position \edt{of the wing}   & $\Delta Y/R$ & $0.50$, $0.75$, $1.00$ \\
Axial position \edt{of the wing}     & $\Delta X/R$ & $0.50$, $0.75$, $1.00$ \\
\hline
\end{tabular}
\end{table}

Before examining the VTOL propeller-wing configuration in hover, the percentage increment in each operating parameter relative to its preceding value is calculated. This approach facilitates the assessment of correlations between variations in the input parameters and the resulting aerodynamic response, \edt{which allows} the associated performance trade-offs to be more clearly interpreted. The increment~($\%$) is calculated according to the relation provided below. 

\begin{eqnarray}
    Increment~(\%)= \left|\frac{X_{i}-X_{{i-1}}}{X_{i}}\right| \times 100\%
\label{eq:deviation}
\end{eqnarray}

\noindent where $X_i$ denotes the current parameter value\edt{,} and $X_{i-1}$ \edt{shows} the preceding value. The calculated increments are listed in Table~\ref{tab:parameter_increment}. The increments are substantial, ranging from approximately $20\%$ to $40\%$, \edt{and provide} a broad parametric framework for evaluating the sensitivity of the aerodynamic response. A comparable variation in the performance metrics indicates a strong influence of the corresponding parameter, whereas a limited response despite the large \edt{parametric} increment \edt{highlight} weak sensitivity and an unfavorable trade-off \edt{in the performance of the aerodynamic design}.

\begin{table}[h!]
\caption{\label{tab:parameter_increment}
Relative increments in the governing input parameters for the propeller-wing configuration in hover.}
\centering
\begin{tabular}{cccccc}
\hline
\textbf{RPM} &
\textbf{Increment (\%)} &
$\boldsymbol{\Delta X/R}$ &
\textbf{Increment (\%)} &
$\boldsymbol{\Delta Y/R}$ &
\textbf{Increment (\%)} \\
\hline

$8,900$  & --        & --     & --        & --     & --        \\
$11,430$ & $22\%$   & $0.50$ & --        & $0.50$ & --        \\
$15,740$ & $37\%$   & $0.75$ & $33\%$   & $0.75$ & $33\%$   \\
$24,200$ & $35\%$   & $1.00$ & $25\%$   & $1.00$ & $25\%$   \\

\hline
\end{tabular}
\end{table}

The parametric framework used to \edt{investigate} the propeller-wing interaction in hover differs from that employed in cruise. The two principal performance metrics considered are the figure of merit ($FM$) and $C_T$ \edt{\citep{zhao2025evolution}}. In \edt{this situation}, $C_T$ does not represent the forward-flight capability of the propeller\edt{. Rather}, it provides a non-dimensional measure of the thrust generated and\edt{,} therefore\edt{,} indicates the lifting capability available for vertical take-off. The figure of merit, commonly used \edt{for quantifying performance} in helicopters and tilt-rotors \edt{\cite{leishman2006principles}}, is a non-dimensional measure of hovering efficiency\edt{. It is} defined by comparing the ideal power required to generate a given thrust with the actual power consumed by the propeller. The figure of merit of small-scale rotors is often below $50\%$ \edt{due to} increased viscous, induced, and profile-\edt{related} power losses \edt{\cite{leishman2006principles}. It} is consistent with the results presented in Fig.~\ref{fig:hover_quantitative}a, where the maximum figure of merit is $FM=42.8\%$.

\begin{figure}[h!]
    \centering
    \includegraphics[width=1\linewidth]{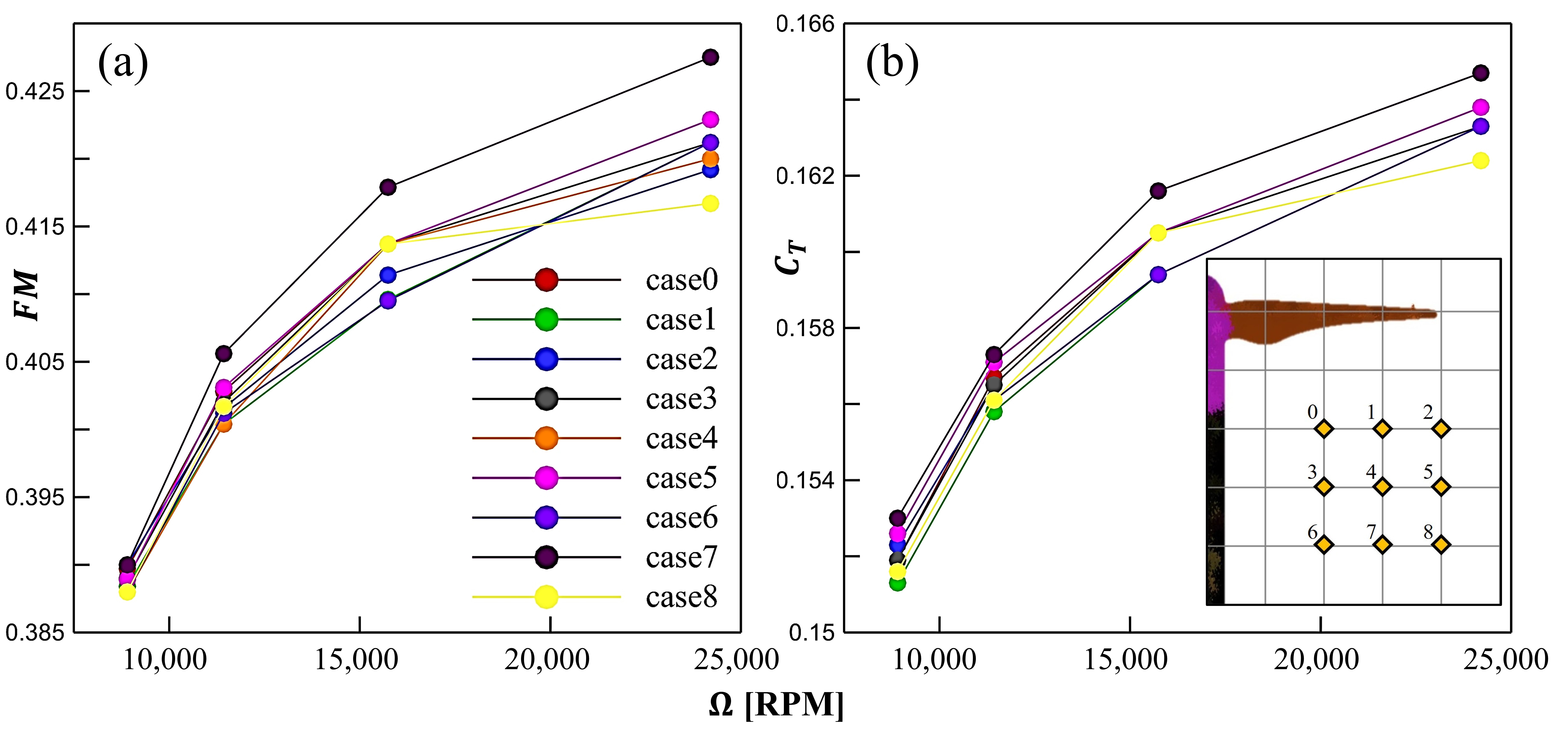}
    \caption{\dn{(a) Figure of merit and (b) Thrust coefficient for the propeller-wing configuration at different rotational speeds and locations of the wing, identified by cases}}
    \label{fig:hover_quantitative}
\end{figure}

Figures~\ref{fig:hover_quantitative}a and \ref{fig:hover_quantitative}b \edt{depict} variations in $FM$ and $C_T$, respectively, with \edt{the} propeller\edt{'s} rotational speed for all positions \edt{of the wing} relative to the propeller\edt{'s} axis of rotation \edt{under consideration here}. The legends in Fig.~\ref{fig:hover_quantitative}a \edt{identify} the case numbers corresponding to the wing\edt{'s location} shown in the inset of Fig.~\ref{fig:hover_quantitative}b. Both plots exhibit similar trends, indicating that the increase in $C_T$ is accompanied by an increase in $FM$. Although both parameters generally increase with $\Omega$, the differences between their minimum and maximum values are only $9.5\%$ for $FM$ and $8\%$ for $C_T$. These variations occur over a $63\%$ increase in rotational speed, from $\Omega=8,900$ to $24,200~\mathrm{RPM}$. This trend reveals two key observations. \edt{Firstly}, the propeller-wing configuration can operate at higher rotational speeds in hover without a significant variation in hovering efficiency, as indicated by the relatively stable $FM$. \edt{Secondly}, the increase in $C_T$ does not scale linearly with $\Omega$. These findings provide important design considerations for small-scale tilt-rotor aircraft operating in hover. Furthermore, the closely clustered curves in Fig.~\ref{fig:hover_quantitative} further \edt{portrays} that the propeller\edt{'s} performance is only weakly influenced by the position \edt{of the wing in its wake during hover}.

To examine the effect of varying the propeller rotational speed, case~8 \edt{with $\Delta X/R=\Delta Y/R=1.0$} is selected, corresponding to the \edt{farthest position of the wing from the propellor}. Figure~\ref{fig:velocity_case8_X} \edt{provides} \edt{contours of the} streamwise velocity \edt{($U_Y$)}\edt{, extracted from downstream of} the propeller\edt{'s} plane. \edt{This component of velocity is} parallel to the propeller\edt{'s} axis of rotation, $U_Y$. The \edt{range of} velocity \edt{is chosen} from $U_Y=0$ to $36~\mathrm{m/s}$. For comparison, the maximum freestream velocity considered in the \edt{our analysis for the cruise condition is} $U_{\infty}=12~\mathrm{m/s}$. Therefore, the upper limit selected for \edt{contours of} the \edt{flow} velocity \edt{in  the hovering conditions} is three times the maximum cruise velocity.

\begin{figure}[h!]
    \centering
    \includegraphics[width=1\linewidth]{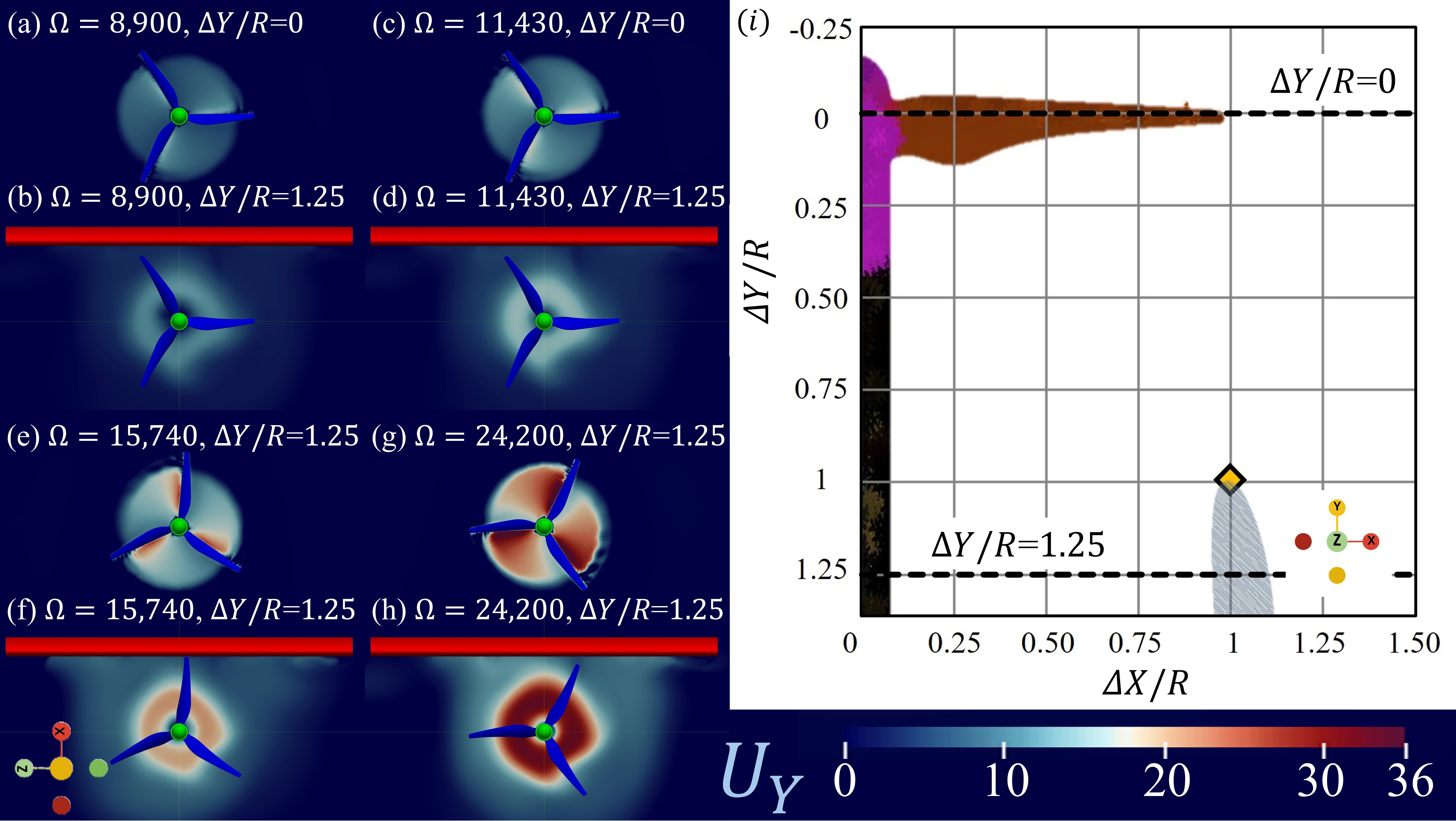}
    \caption{\edt{Contours of ($U_Y$)} for case~8 at (a \& b) $\Omega=8,900~\mathrm{RPM}$, (c \& d) $\Omega=11,430~\mathrm{RPM}$, (e \& f) $\Omega=15,740~\mathrm{RPM}$, and (g \& h) $\Omega=24,200~\mathrm{RPM}$. The contours are extracted at cross-sectional planes located at $\Delta Y/R=0$ and $\Delta Y/R=1.25$, respectively. (i) The schematic here \edt{shows} the locations of the two sampling planes relative to the propeller and wing.}
    \label{fig:velocity_case8_X}
\end{figure}

The velocity contours shown in \edt{Fig.~\ref{fig:velocity_case8_X}} are extracted at two planes parallel to the $X$--$Z$ plane, located at $\Delta Y/R=0$ and farther downstream at $\Delta Y/R=1.25$. \edt{Figures}~\ref{fig:velocity_case8_X}a and \ref{fig:velocity_case8_X}b correspond to $\Omega=8,900~\mathrm{RPM}$ at $\Delta Y/R=0$ and $\Delta Y/R=1.25$, respectively. Similarly, Figs.~\ref{fig:velocity_case8_X}c and \ref{fig:velocity_case8_X}d correspond to $\Omega=11,430~\mathrm{RPM}$, Figs.~\ref{fig:velocity_case8_X}e and \ref{fig:velocity_case8_X}f correspond to $\Omega=15,740~\mathrm{RPM}$, and Figs.~\ref{fig:velocity_case8_X}g and \ref{fig:velocity_case8_X}h correspond to $\Omega=24,200~\mathrm{RPM}$. At $8,900~\mathrm{RPM}$, the propeller generates a jet-like axial velocity of approximately $U_Y=18~\mathrm{m/s}$, which exceeds the freestream velocity of $U_{\infty}=8~\mathrm{m/s}$ considered at $J=0.8$ in the cruise configuration \edt{elucidated} in Fig.~\ref{fig:framework}. An important distinction between the cruise and hover conditions is that increasing the rotational speed at a fixed cruise velocity reduces $J$ and $\eta$, whereas $FM$ in hover increases over the investigated \edt{range of} rotational speed \edt{of the propellor (see} Figs.~\ref{fig:eff} and \ref{fig:hover_quantitative}\edt{)}. \edt{Contours of} the axial velocity indicate a substantial increase in slipstream momentum with \edt{an} increasing rotational speed. At $\Omega=24,200~\mathrm{RPM}$, Figs.~\ref{fig:velocity_case8_X}g and \ref{fig:velocity_case8_X}h \edt{exhibit} that the high-velocity region becomes concentrated near the center \edt{of the wake}. The persistence of this concentrated velocity core at $\Delta Y/R=1.25$ indicates that the slipstream retains substantial jet-like axial momentum farther downstream.

\begin{figure}[h!]
    \centering
    \includegraphics[width=1\linewidth]{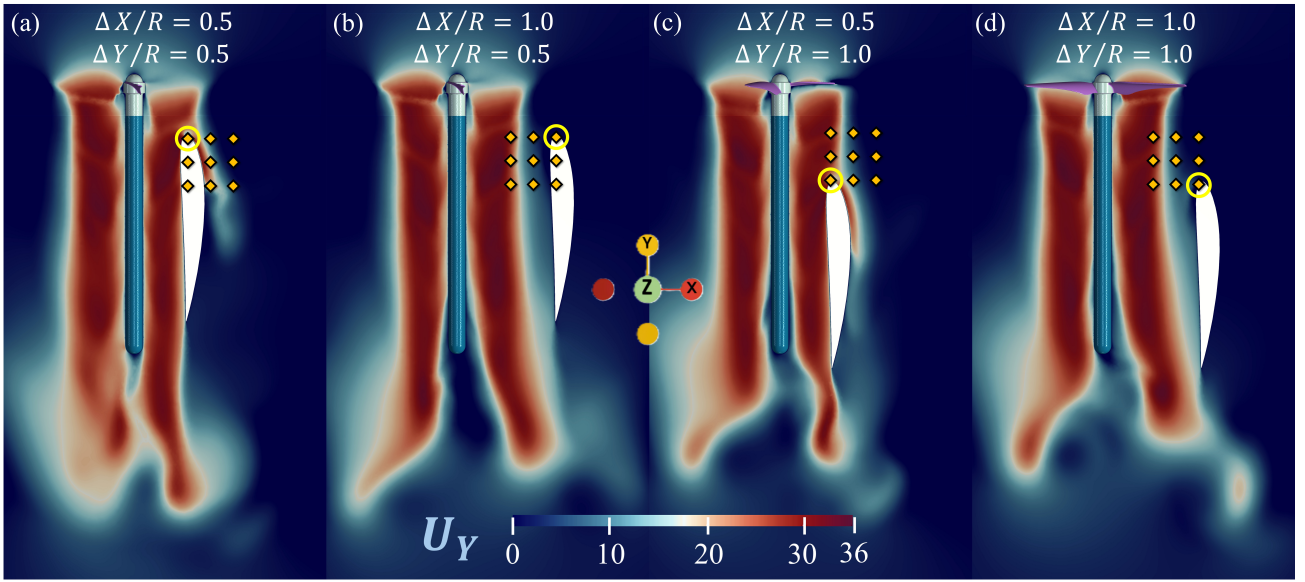}
    \caption{\edt{Contours of $U_Y$ from the side view} for the propeller-wing configuration in hover at $\Omega=24,200~\mathrm{RPM}$ for \edt{different} positions \edt{of the wing} (a) $\Delta X/R=0.50$ \edt{and} $\Delta Y/R=0.50$\edt{,} (b) $\Delta X/R=1.00$ \edt{and} $\Delta Y/R=0.50$\edt{,} (c) $\Delta X/R=0.50$ \edt{and} $\Delta Y/R=1.00$, and (d) $\Delta X/R=1.00$ \edt{and} $\Delta Y/R=1.00$. \edt{Here,} the circled markers identify the corresponding wing's locations.}
    \label{fig:velocity_side_Hover_24200}
\end{figure}

Although the quantitative analysis presented earlier \edt{hints} that the wing\edt{'s} position has very little influence on the overall performance of the \edt{aerodynamic configuration here}, its four \edt{different placements} are shown in Figs.~\ref{fig:velocity_side_Hover_24200}a-\ref{fig:velocity_side_Hover_24200}d. In each contour, the wing\edt{'s} location is identified using the markers introduced in Fig.~\ref{fig:wingplacement}b to facilitate \edt{the} comparison \edt{of flow fields in these plots. Here, we observe} two key features \edt{the two cases in} Figs.~\ref{fig:velocity_side_Hover_24200}a and \ref{fig:velocity_side_Hover_24200}c ($\Delta X/R=0.5$)\edt{: (1)} the flow is partitioned at the leading edge of the wing, and this bifurcation is visible in both \edt{contour plots}. Although this interaction may introduce additional aerodynamic losses, the earlier quantitative assessment indicates that its overall influence is limited, as the corresponding variation in drag is not significant\edt{, and (2)} compared with their counterparts shifted farther in the $X$-direction, as shown in Figs.~\ref{fig:velocity_side_Hover_24200}b and \ref{fig:velocity_side_Hover_24200}d, the configurations with the wing closer to the nacelle produce a more pronounced jet-like velocity field that persists farther downstream in the streamwise direction. Another important phenomenon observed in Figs.~\ref{fig:velocity_side_Hover_24200}b and \ref{fig:velocity_side_Hover_24200}d differs from the flow impingement \edt{on the leading edge of the wing} discussed previously. When the wing is positioned farther downstream, a larger portion of the high-velocity flow passes beneath the wing rather than over its upper surface. \edt{On the contrary}, the pressure difference between the pressure side above the wing and the suction side beneath the wing promotes the bifurcation of the slipstream around the wing at its trailing edge.

\section{Conclusions}

The present study investigates the aerodynamic \edt{propellor-wing} interactions \edt{in} VTOL \edt{UAVs} under cruise and hover conditions. \edt{For the} cruise \edt{condition}, the present study connects these quantitative results with the underlying wake dynamics, including flow impingement, vortex stretching, slipstream bifurcation, and downstream vortex convection. The maximum thrust coefficient under cruise conditions is observed at lower values of $J$ ($0.2$-$0.3$). \edt{On the other hand}, the propulsive efficiency generally increases with $J$ and reaches its maximum near $J=0.6$-$0.7$. The highest efficiency, $\eta=61\%$, is obtained at $J=0.7$ and $U_{\infty}=12~\mathrm{m/s}$. At higher $J$, the propeller generates a narrower and more coherent slipstream that retains its jet-like axial momentum farther downstream after interacting with the wing. The associated wake evolution is characterized through three phases: the approaching phase, interaction phase, and convection phase. \edt{For hovering flights}, the aerodynamic performance is evaluated using $FM$ and $C_T$ for over a range of wing\edt{'s locations} and propeller\edt{'s} rotational speeds. Our results show that variations in the axial and vertical positions \edt{of the wing} have only a limited influence on the overall $C_T$ and $FM$. In comparison, increasing the $\Omega$ produces an overall increase in both performance parameters, although the aerodynamic gains do not scale proportionally. The maximum $FM$ obtained is $0.428$. The flow-field analysis further demonstrates that increasing $\Omega$ strengthens the concentrated jet-like slipstream and allows substantial axial momentum to persist farther downstream.


\section*{Acknowledgment}
MSU Khalid acknowledges the funding support from the Natural Sciences and Engineering Research Council of Canada (NSERC) through the Discovery grant program. S. Choi and MSU Khalid are thankful to MITACS for their support through the Globalink Research Internship (GRI) program. The simulations reported in this work were performed on the supercomputing clusters administered and managed by the Digital Research Alliance of Canada. 


\vskip6pt

\bibliography{references}

\end{document}